%% file: Maximum_Likelihood_Spectrum_Decomposition_for_Isotope_Identification_and_Quantification.tex
\documentclass[
10pt,
final,
journal,
letterpaper,
oneside,
twocolumn
]{IEEEtran}

\usepackage{cite}                     % better citation handling
\usepackage{textcomp}                 % more fonts
\usepackage{nicefrac}                 % nicer fraction typesetting
\usepackage{graphicx}                 % Include figure files
\usepackage{bm}                       % bold math
\usepackage{amsmath,amssymb,amsfonts} % for the align environment and other nice math stuff like case environments
\usepackage{algorithm}                % for algorithm blocks
\usepackage{algcompatible}            % for algorithm blocks
\usepackage{algpseudocode}            % for algorithm blocks
\usepackage{hyperref}                 % make references become links
\usepackage{xfrac}                    % somewhat better looking ``swing fractions''
\usepackage{lscape}                   % to get the landscape table for the comparison table

\newcounter{DblColEqnCounterTemp}

\begin{document}

    \bstctlcite{IEEEexample:BSTcontrol}

    \title{Maximum Likelihood Spectrum Decomposition for Isotope Identification and Quantification}
    \author{J. T. Matta, A. J. Rowe, M. P. Dion, M. J. Willis, A. D. Nicholson, D. E. Archer, H. H. Wightman
    \thanks{This research was supported by the U.S. Department of Energy, National Nuclear Security Administration, Office of Defense Nuclear Nonproliferation Research and Development (DNN R\&D) and was performed at Oak Ridge National Laboratory managed by UT-Battelle LLC for the U.S. Department of Energy under Contract DE-AC05-00OR22725.

    This manuscript has been co-authored by UT-Battelle, LLC, under contract DE-AC05-00OR22725 with the US Department of Energy (DOE). The US government retains and the publisher, by accepting the article for publication, acknowledges that the US government retains a nonexclusive, paid-up, irrevocable, worldwide license to publish or reproduce the published form of this manuscript, or allow others to do so, for US government purposes. DOE will provide public access to these results of federally sponsored research in accordance with the DOE Public Access Plan (http://energy.gov/downloads/doe-public-access-plan).}
    \thanks{J. T. Matta (email: mattajt@ornl.gov) and A. J. Rowe are with the Science and Technology Software Group, Cadre5, LLC, Knoxville, TN 37932, USA}
    \thanks{M. P. Dion is with the Nuclear Nonproliferation Division, National Security Sciences Directorate, Oak Ridge National Laboratory, Oak Ridge, TN, 37831, USA}
    \thanks{M. J. Willis, A. D. Nicholson, and D. E. Archer are with the Physics Division, Physical Sciences Directorate, Oak Ridge National Laboratory, Oak Ridge, TN, 37831, USA}
    \thanks{H. H. Wightman is with the Chemical Sciences Division, Physical Sciences Directorate, Oak Ridge National Laboratory, Oak Ridge, TN, 37831, USA}}

    \maketitle

    \begin{abstract}
        A spectral decomposition method has been implemented to identify and quantify isotopic source terms in high-resolution gamma-ray spectroscopy in static geometry and shielding scenarios.
        Monte-Carlo simulations were used to build the response matrix of a shielded high purity germanium detector monitoring an effluent stream with a Marinelli configuration.
        The decomposition technique was applied to a series of calibration spectra taken with the detector using a multi-nuclide standard.
        These results are compared to decay corrected values from the calibration certificate.
        For most nuclei in the standard ($^{241}$Am, $^{109}$Cd, $^{137}$Cs, and $^{60}$Co) the deviations from the certificate values were generally no more than $6$\% with a few outliers as high as $10$\%.
        For $^{57}$Co, the radionuclide with the lowest activity, the deviations from the standard reached as high as $25$\%, driven by the meager statistics in the calibration spectra.
        Additionally, a complete treatment of error propagation for the technique is presented.
    \end{abstract}

    \section{Introduction}
    \input{intro}

    \section{Theory}
    \input{theory}

    \section{Instrument Setup}
    \input{instrument_setup}

    \section{Response Matrix Generation}
    \input{resp_mat_gen}

    \section{Analysis}
    \input{analysis}

    \section{Results}
    \input{results}

    \section{Conclusions and Future Improvements}
    \input{conclusion}

    \appendix
%    \appendices
    \section{Propagation of Uncertainties}
    \input{appendix_a}

    \section*{Acknowledgment}
        The authors would like to thank Marc Chattin of the Radioactive Materials Analytical Lab in the Chemical Sciences Division of the ORNL Physical Sciences Directorate for maintaining the detector and its data acquisition software, execution of calibration runs, and willingness to answer questions about and produce documentation on the effluent detector.
        The authors would also like to thank Michael Stafford, division director of the Nuclear \& Radiological Protection Division of the ORNL Environmental, Safety, Health, and Quality Directorate, for permission to use the data collected by the effluent monitor for this research.
        Finally, the authors would like to thank the staff and researchers of HFIR and the REDC for tolerating our intrusions and answering our frequent questions about reactor operations and isotope production campaigns.

%    \section*{References}
    \bibliographystyle{IEEEtran}
    \bibliography{references}

\end{document}

%% file: intro.tex
    Gamma-ray detectors are highly versatile systems used for, among other things, monitoring nuclear facilities and remote data collection.
    One such scenario is the measurement of effluent streams to identify and quantify aerosol and gaseous radionuclides.
    Some examples of these monitoring scenarios are found at nuclear power reactors \cite{NRC_effluent}, medical isotope production facilities \cite{DECONNINCK201361}, and other nuclear research institutions.
    The High Flux Isotope Reactor (HFIR) at the Oak Ridge National Laboratory (ORNL) is a research reactor designed to produce extremely high fluxes of thermal and cold neutrons.
    It is a potent tool for the scientific community enabling neutron scattering, isotope production, materials irradiation testing, and neutron activation analysis.
    Adjacent to HFIR is the Radiochemical Engineering Development Center (REDC), where various chemical processes are performed for isotope production campaigns.
    These facilities share a common effluent stack and the ORNL is required to monitor their emissions from this stack and publish the results annually\cite{EPA_effluent}.

    The effluent from the common stack of the HFIR campus is monitored using a flow-through Marinelli beaker with a high purity germanium (HPGe) detector encased in an environmental shield to reduce background radiation.
    The detector is calibrated weekly with a multi-nuclide source constructed to approximate the solid angle and efficiency profile of the Marinelli beaker.
    Specific nuclide release amounts must be monitored accurately to adhere to the requirements in 40 CFR Part 61, Subpart H\cite{EPA_effluent} - National Emission Standards for Hazardous Air Pollutants (NESHAPS).
    The spectral analysis for the environmental monitoring uses a peak fitting technique with commercial software (Ortec GammaVision\cite{ortecGammaVision}).
    The method is applied to conform to the reporting policies and the data stream and analysis cannot be interrupted.

    In applications where a detector system with static geometry and shielding is used to identify and quantify isotopes in an effluent stream continuously, a method that provides accurate results with little intervention is highly desirable.
    This research presents the details and validation of a maximum likelihood spectral decomposition technique that fulfills these requirements when supplied with response functions determined through high-fidelity particle transport simulations.
    In addition, since a parallel data stream is analyzed (for reporting purposes) using commercial software that has been validated, these results serve as a way to compare the results from the decomposition technique on high energy resolution gamma-ray spectra.
    This technique provides the absolute number of radioisotope decays over the time of spectrum integration and is validated by decomposing spectra measured by the HFIR campus common stack effluent detector during its weekly calibrations.
    The decomposition results for the activities of the radionuclides in the calibration standard are in good agreement with the decay corrected quantities calculated from the source calibration certificate.

%% file: theory.tex
\subsection{Solving the Inverse Problem}
    In its most basic form, spectral decomposition is a standard, if complex, inverse problem.
    As applied to spectral response, the inverse problem is to find the input terms (nuclide source terms in this work) that result in the measured detector spectrum using the detector responses encoded in the response matrix.
    Frequently these problems can be written as a simple matrix equation:
    \begin{align}
        \label{eqn:basic_inverse_eqn}
        \phantom{.}\mathbf{R}\vec{I}=\vec{O}.
    \end{align}
    Where $\mathbf{R}$ is the response matrix of the detector (obtained by measurement or simulation,) $\vec{I}$ is the input to the detector (e.g., decays of isotopes, number of gamma rays at a particular energy,) and $\vec{O}$ is the observed spectrum.

    When the problem is written as in \ref{eqn:basic_inverse_eqn} a straightforward solution presents itself.
    Simply find $\mathbf{R}^{-1}$ (or, if $\mathbf{R}$ is not square, the Moore-Penrose inverse $\mathbf{R}^{+}$) and one can obtain $\vec{I} = \mathbf{R}^{-1} \vec{O}$.
    However, there are problems with the matrix inversion approach.
    The first is that to be invertible, a matrix must not be singular (i.e., not have a determinant of zero), equivalent to requiring that the rows be linearly independent.
    For response matrices, linear dependence of rows should be a rare problem; but it can occur depending on the inputs used to produce the matrix.
    The second problem is that response matrices are frequently ill-conditioned.
    In an ill-conditioned matrix the rows of the response matrix are, mathematically, linearly independent but the changes to the coefficients required to make the matrix singular are small.
    In such cases, many inversion procedures will experience a build-up of floating-point arithmetic errors and produce a matrix that will \emph{not} yield the identity matrix when multiplied by the original matrix.
    The final problem is that it is common for some of the coefficients of $\vec{I}$ to have negative values.
    While the coefficient of an input to a spectrum can be zero, it cannot ever be negative as that would be equivalent to removing energy depositions from the detector; a nonphysical notion.

    The need to solve the spectral inverse problem comes up in many fields.
    The maximum likelihood expectation maximization (MLEM) method was first developed for statistical astronomy by L.B. Lucy\cite{lucyAstro1974}.
    A more rigorous derivation of its properties was later given in \cite{dempster1977}.
    Later it was applied to emission and transmission tomography in \cite{tomography1, tomography2}.
    More recently, MLMEM has been applied to gamma-ray total absorption spectroscopy\cite{tainOtt2007, rascoMTAS}; which motivated this research effort.

    The MLEM approach is immune to the problems of singular and ill-conditioned matrices.
    Additionally, MLEM maximizes each bin's Poisson likelihood, appropriate for counting-statistics-driven data like spectra.
    Further, if there are fewer response functions than spectrum bins, the value extracted is unique\cite{lucyAstro1974}.
    Finally, weights produced will be positive, removing the possibility of nonphysical solutions.

    The iterative equation used in this research is given in \ref{eqn:decomp_eqn}.
    Some care should be taken in examining \ref{eqn:decomp_eqn}, the individual response functions are placed on rows instead of columns, making it the transpose of the standard definition.

    \begin{align}
        \label{eqn:decomp_eqn}
        I_{\mu{}}^{s+1} = \frac{1}{\sum_{j}R_{\mu{}j}}\sum_{i}\frac{I_{\mu{}}^{s}R_{\mu{}i}O_i}{\sum_{\alpha{}}R_{\alpha{}i}I_{\alpha{}}^{s}}
    \end{align}
    Here $\mu$ and $\alpha$ are row indices; $j$ and $i$ are column indices; and $R$, $I$, and $O$ have the same meanings as in \ref{eqn:basic_inverse_eqn}, but specific coefficients are being referenced instead of the whole matrix or vector.

    The iterative MLEM method requires an initial, non-zero, guess vector $\vec{I}^{(s=0)}$ and each iteration refines the values.
    Iteration stops when the coefficients of $\vec{I}$ have all either passed below a consideration threshold or converged.
    The consideration threshold is a defense against the finite precision of computer floating-point numbers.
    Coefficients less than this threshold are ignored for testing convergence; because negligibly small values can have significant proportional changes while still retaining a negligible magnitude.
    Parameters above the consideration threshold are judged converged when the magnitude of their change relative to their previous value is less than some threshold.
    For convergence threshold $T$ and consideration threshold $C$ the iterative MLEM has converged when $\sfrac{|I^{(s)}_\mu - I^{(s-1)}_\mu|}{I^{(s-1)}_\mu} \le T$ for all $I^{(s)}_\mu > C$.
    The Appendix describes two modes of error propagation that can be carried out post-convergence.

\subsection{Response Matrices}

    The response matrix is an essential construct in this work.
    Each row in the matrix is a response function.
    A response function describes the spectrum observed in response to a particular input.
    These inputs could be individual specific energies of mono-energetic photons, the average emission from the decay of a single nucleus, etc.
    The choice depends on the desired output of the system.
    After decomposition, the interpretation of the weights obtained for each row is determined by what those rows represent.

    The construction of a response matrix can proceed down two parallel paths.
    The matrix can be measured in a series of experiments or simulated in a particle transport code.
    Measurement of the response matrix poses many challenges.
    For instance, obtaining intense sources of mono-energetic photons can be problematic, as is obtaining sufficiently large and pure quantities of any number of short-lived isotopes.
    Further, if the detector's electronic noise level changes, the detector accumulates radiation damage, etc., the peak widths will change, necessitating the remeasurement of the entire response matrix.

    Performing high-fidelity simulation presents issues of its own when applied exclusively.
    The geometry used in a simulation is rarely a perfect rendition of the detector and shielding.
    Sometimes simplifying approximations must be made and, many times, there is information that is not known or available.
    Nor can simulation easily account for minor imperfections in the detector or electronics.
    Imperfections which add noise, change the active volume, or affect charge/light collection are difficult to quantify.
    Nor can the simulation account for event to event variations in charge/light collection.
    These variations produce differences in pulse height that cause a level threshold to suppress counts below some energy probabilistically.
    Finally, simulating the charge/light collection and propagation through the detector to derive peak widths requires immense effort to generate accurate results.

    Thankfully, a hybrid approach suggests itself; simulation extracts energy deposition probabilities while other aspects of the detector response are determined empirically and applied \emph{post facto}.
    Minor inaccuracies in geometry can be handled by appropriate scaling of the response functions to account for missing or too thin absorbing layers.
    Charge/light collection and electronic noise are folded into the response by convolving the energy deposition histogram with a Gaussian function whose width depends on some empirically determined function of deposited energy.
    Changes in the detector resolution are compensated for by simply recomputing the convolution with updated width parameters, taking a few minutes at most.
    Threshold effects can be dealt with either by truncating the response function at an energy just above where the effects disappear or as some empirically determined probabilistic culling of counts close to the threshold.

    In this work, the hybrid approach was employed.
    Energy deposition response functions were simulated in Geant4 10.05.p01 \cite{Geant4Ref1, Geant4Ref2, Geant4Ref3} using the data libraries it downloaded in the build process.
    These were convolved with a realistic detector width (discussed in greater detail later) and threshold effects were dealt with by, prior to decomposition, truncating the response functions and input spectra below ${\scriptstyle \sim}0.045$ MeV.
    With this approach, response functions that match reality were produced with a smaller investment of human effort and computation time than would otherwise be required to produce response functions of similar fidelity in a purely \emph{ab initio} manner.

%% file: instrument_setup.tex
\begin{figure}
    \includegraphics[width=\columnwidth, trim=10cm 0 7cm 0, clip]{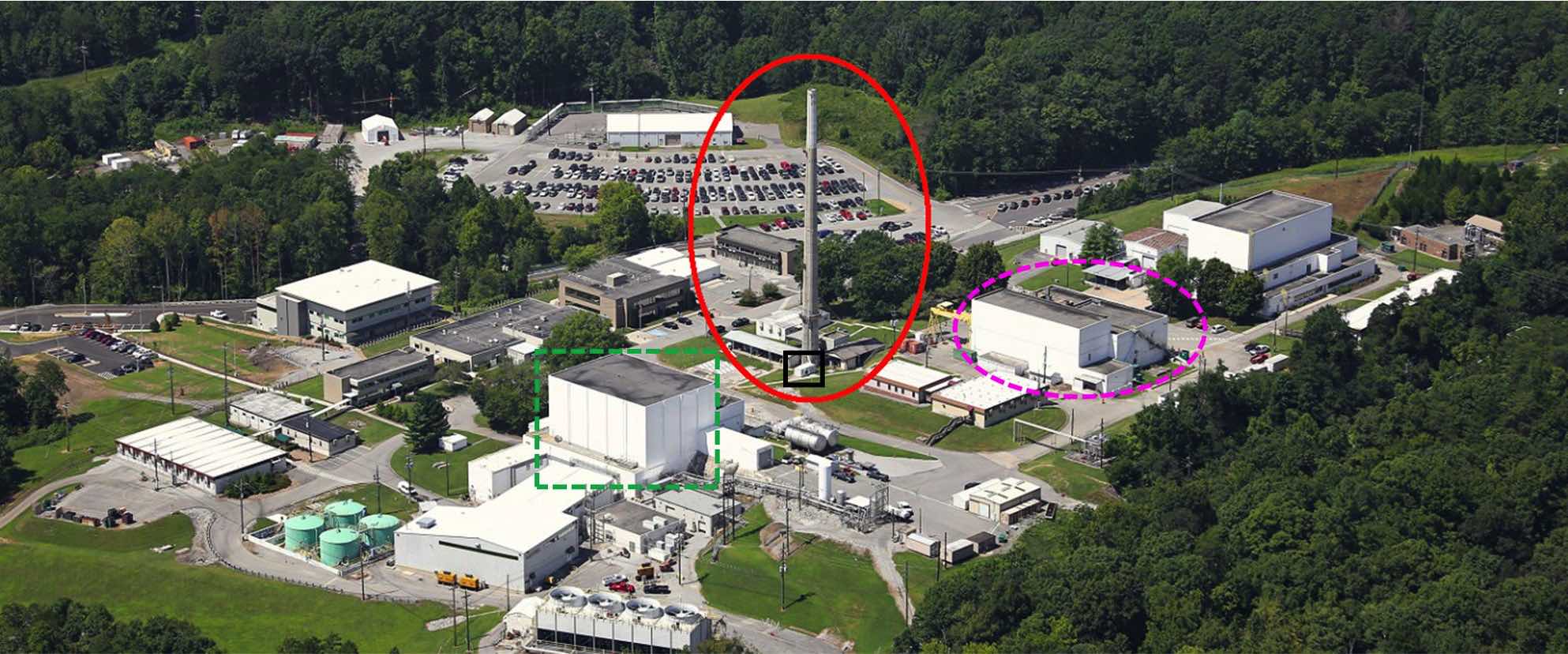}
    \caption{\emph{Color Online.} An aerial image of the ORNL campus showing the HFIR facility (dashed green box), main stack (solid red oval), the REDC (dashed magenta oval), and the small building that houses the effluent detector (solid black square).\label{fig:hfir-picture}}
\end{figure}

\begin{figure}
    \includegraphics[width=\columnwidth]{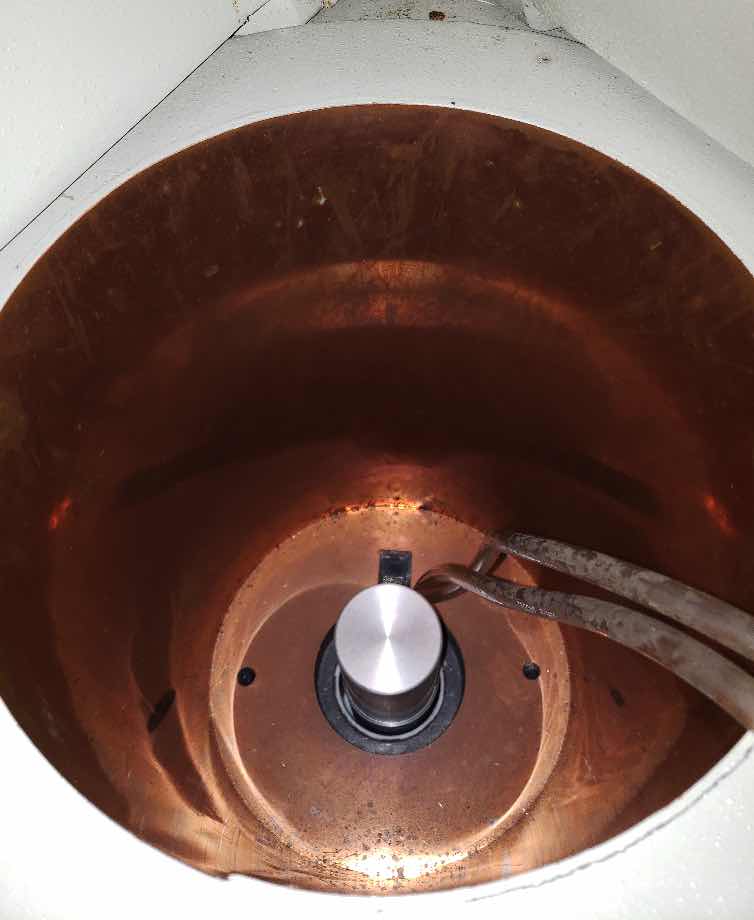}
    \caption{\emph{Color Online.} An image of the interior of the detector setup.
        The Marinelli beaker has been temporarily removed prior to inserting the calibration source, but the gas lines that flow to it can be seen.\label{fig:interior-det-setup}}
\end{figure}

\begin{figure}
    \includegraphics[width=\columnwidth]{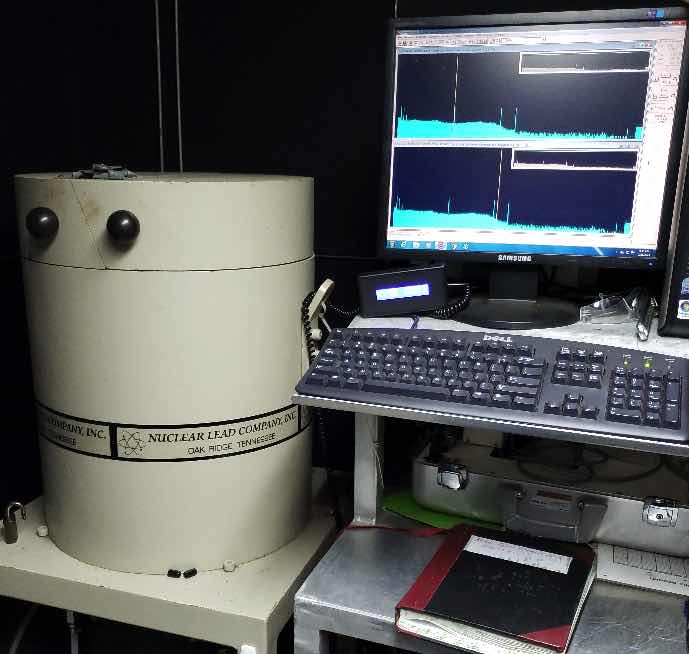}
    \caption{\emph{Color Online.} An image of the exterior of the detector setup.
        The shield on its stand is shown to the left of the computer. \label{fig:exterior-det-setup}}
\end{figure}

\begin{table*}
    \caption{\label{tbl:paraphrased-source-certificate}A table of source data from or derived from the detector calibration source's certificate and the Evaluated Nuclear Structure Data Files (ENSDF)\cite{ensdfDbReport}.
        The Primary $\gamma$s column shows only the gamma rays listed on the source certificate.
        Half life and gamma-ray intensities are drawn from the ENSDF, their uncertainties are $1\sigma$.
        The ``Initial Activity'' value was calculated from the ``$\gamma$ps'' listed on the certificate with a $1\sigma$ error of $\sfrac{1}{2}$ the ``Expanded Relative Uncertainty'' (which is given for k=2) given in the certificate and the intensity of the principal gamma ray(s) with both errors propagated appropriately. }
    \begin{center}
        \begin{tabular}{c||c|c|c|c}
            \hline
            Nuclide    & Half-Life        & Primary $\gamma$s [MeV] & $\gamma$ Intensity     & Initial Activity [Bq] \\
            \hline
            $^{241}$Am & $432.6   (6)$ y  & $0.0595$             & $0.359   (4)$                  & $1971.0  (417)$ \\
            \hline
            $^{109}$Cd & $461.4   (12)$ d & $0.0880$             & $0.03664 (16)$                 & $26215.7 (5496)$ \\
            \hline
            $^{57}$Co  & $271.74  (6)$ d  & $0.1221$             & $0.8560  (17)$                 & $607.9   (107)$ \\
            \hline
            $^{137}$Cs & $30.08   (9)$ y  & $0.6617$             & $0.8510  (20)$                 & $767.8 (155)$ \\
            \hline
            $^{60}$Co  & $1925.28 (14)$ d & $1.1732$ \& $1.3325$ & $0.9985 (3)$ \& $0.999826 (6)$ & $1237.5 (171)$ \\
            \hline
        \end{tabular}
    \end{center}
\end{table*}

    The off-gas effluent from HFIR and the REDC is routed through two $1.22$ m diameter steel pipes attached to the base of a $76.2$ m stack (the main stack) and discharged to the atmosphere.
    The effluent stream from the HFIR is filtered for particulates and radioiodine.
    The REDC line also contains mitigation for radioiodine.
    The piping from each facility is designed to handle and discharge ${\scriptstyle \sim}14\times{}10^3~\sfrac{\text{L}}{\text{s}}$ into the main stack which is designed for up to $31\times{}10^3~\sfrac{\text{L}}{\text{s}}$.
    The main stack is situated approximately in between the two facilities as shown in Fig. \ref{fig:hfir-picture}.
    Effluent from the main stack is sampled through a shrouded probe at the height of $15.24$ m from ground level.
    A stainless steel sample line brings the effluent to a ground-level building that houses radiometric counting instrumentation, including the shielded HPGe detector investigated in this research.
    The sampled effluent passes through additional filtration in the building, including a high-efficiency filter paper and three activated charcoal sample cartridges.
    The additional filtration removes particulates and any additional radioiodine produced from radioactive decay and incomplete removal at the source (i.e., the HFIR or REDC) to avoid contamination of the Marinelli beaker.
    The Canberra GC2518 HPGe detector\cite{mirionSegeDet} used in this research is shown within its background shielding setup in Fig. \ref{fig:interior-det-setup}.
    The gas flow is managed by a flow control valve to ${\scriptstyle \sim}0.5~\sfrac{\text{L}}{\text{s}}$ then heated and passed through adsorbent silica gel for tritium removal before entering a $1.2$ L polystyrene Marinelli beaker (GA-MA \& Associates, Inc model G-130G\cite{beakerGAMA}) through plastic tubing at the base of the detector shield (the exterior of which is shown in Fig. \ref{fig:exterior-det-setup} next to the data acquisition computer).
    The effluent is continuously passed through the beaker and returned to the main stack using appropriate pumps.
    The assumption is that the sampling process maintains the radioisotopes' concentrations in the off-gas.
    Therefore, only minimal further effort is required to calculate the total effluent released from the main stack while only measuring a small quantity through sampling.

    The detector is calibrated, approximately weekly, using an Eckert \& Ziegler ``Simulated Gas in 130G GA-MA Gas Beaker'' source\cite{sourceEZ} with reference date: 2015-01-01 12:00 pm.
    This source contained a number of isotopes dispersed evenly through a low density matrix ($0.02~\sfrac{\text{g}}{\text{cm}^3}$).
    The source calibration certificate itself lists not activities but ``$\gamma$s per second'' ($\gamma$ps) as well as the uncertainty of that number.
    Using these values and the ENSDF data for the nuclides, the activities and errors were derived for each gamma ray listed.
    In the case of more than one gamma ray being listed for a single isotope, the activities were combined in an error-weighted average and the errors were combined accordingly.
    Of these isotopes, those that still have the activity to be above the minimum detectable activity (MDA) for moderate calibration durations are found in Table \ref{tbl:paraphrased-source-certificate}.

%% file: resp_mat_gen.tex
    \subsection{Simulation}
\begin{figure}
    \includegraphics[width=\columnwidth, trim=60 100 60 125, clip]{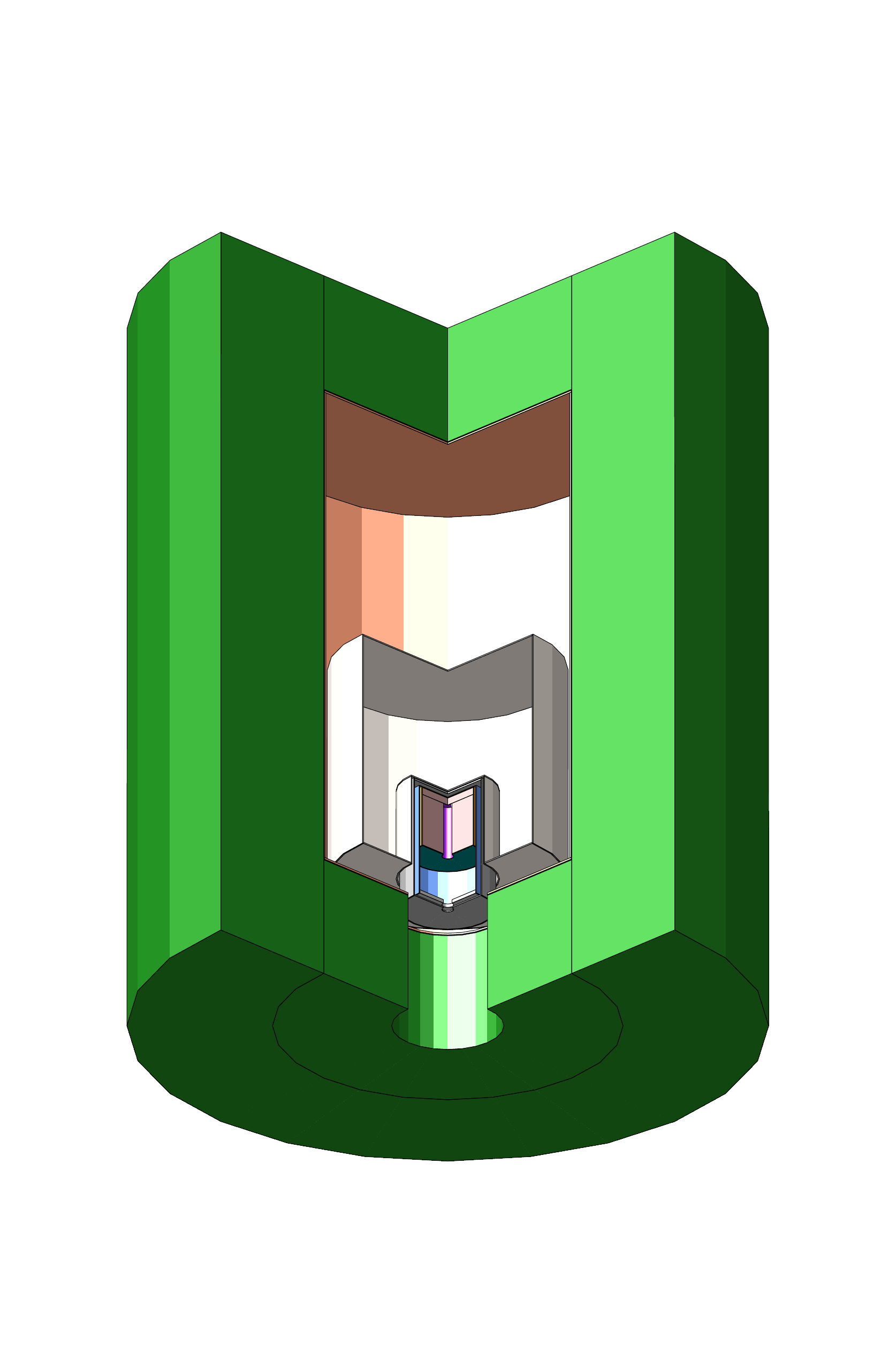}
    \caption{\emph{Color Online.} Visualization of the detector geometry.
        The three layers of the lead shield are visible, with the lead as green, the tin as silver, and the copper as salmon.
        The Marinelli beaker's plastic is rendered in white, and the outer detector cap is rendered in silver.
        The detector's crystal holder sides are light blue, and the aluminized Mylar top and aluminum bottom are gray.
        The dead layer of germanium in the detector's outer contact is orange, the amorphous non-conducting dead layer on the bottom that separates the inner and outer contacts is teal, and the dead layer of the inner contact is purple.
        Finally, the active part of the germanium crystal is the light reddish-brown.
        The shield has an outer radius of $254$ mm, an inner radius of $136.48$ mm, a top and bottom thickness of $101.6$ mm, a copper layer thickness of $1.62$ mm, a tin layer thickness of $0.65$ mm, and a detector entry hole radius of $44.45$ mm.
        The Marinelli beaker has an outer radius of $95.405$ mm, an inner radius of $39.945$, a hollow height of  $76.41$ mm, and a material thickness of $1.5$ mm.
        The detector's outer housing had an outer radius of $38.1$ mm, a sidewall thickness of $1.0$ mm, a window thickness of $0.5$ mm, and a length of $102.5$ mm.
        The detector crystal had an outer radius of $29.15$ mm, a length of $51.6$ mm, and an outer contact dead-layer thickness of $0.7$ mm.
        \label{fig:detector-geometry}}
\end{figure}
    A Monte-Carlo diagram (a simplified engineering diagram) for the detector was requested from Canberra, an engineering diagram of the G-130G was requested from GA-MA \& Associates, and careful measurements were taken of the shield to obtain overall dimensions and layer thicknesses in order to construct a high fidelity simulation.
    The diagrams provided by the manufacturers and the shield measurements were then translated into the Geometry Description Markup Language (GDML)\cite{GdmlRef1} to produce a set of files that Geant4 could ingest to build the geometry.

    The physics list, which provides the details of the interactions used for the simulation, was customized for this research.
    The customized physics list follows the Geant4 standard physics list \emph{QGSP\_BERT\_HP}.
    However, the customized version replaces the less precise, but computationally faster, \emph{G4EmStandard} electromagnetic interactions with the \emph{G4EmLivermore} electromagnetic interactions which produce results that better match observations for low energy ($\lesssim 100$ MeV) electromagnetic interactions\cite{livermorePhysics, livermorePhysicsImprovements}.

    The engineering drawings, while helpful, were still lacking some geometric information.
    Details of the cold finger and the pre-amplification electronics were missing, so they were not included in the simulated geometry.
    However, as neither the cold-finger nor electronics are in the direct path of most of the photons exiting the Marinelli Beaker, it was judged that they were unlikely to be significant scatter sources.
    The crystal cylinder's top edge fillets and the fillets at the top of the hole in the crystal center (where the cold finger and one of the electrical contacts would go) were also omitted.
    The material quantities these fillets contribute or remove are minimal, and their edge placement should minimize their impact on the bulk.
    The final omission is the plastic tubing connected to the Marinelli beaker.
    Due to the tubing's low density and small atomic number, its impact should be minimal, fortunate because the tubing is routinely shifted during calibrations.
    The simulated geometry constructed for this research is shown in Fig. \ref{fig:detector-geometry}.

\subsubsection{Foreground Energy Simulation}
\begin{figure}
    \includegraphics[width=\columnwidth]{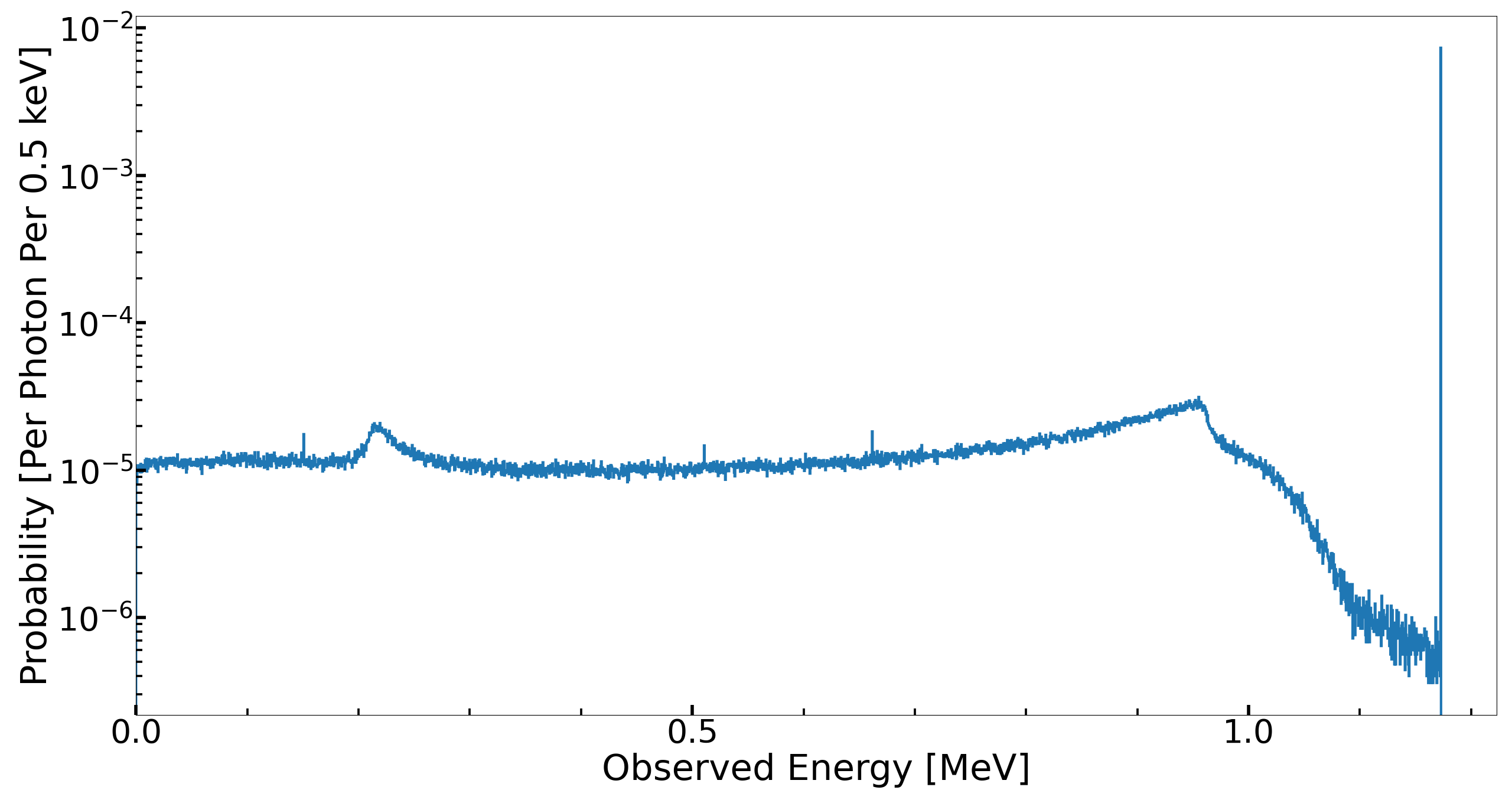}
    \caption{Energy deposition for the 1.173 MeV gamma ray of $^{60}$Co sourced from within the beaker. The single and double escape peaks from pair production and annihilation within the active region of the crystal are visible at $0.662$ MeV and $0.151$ MeV, respectively. The annihilation photon line from pair production outside the active region of the crystal is visible at $0.511$ MeV. The broad peak-like structure around $0.210$ MeV is due to photons that have undergone $180^\circ$ Compton scattering of photons outside the detector and subsequently been absorbed. The shoulder visible around $0.960$ MeV is the Compton edge, produced by photons depositing the maximum energy possible in a single Compton scatter before escaping the detector.\label{fig:en-resp-bkr}}
\end{figure}
    The foreground energy response generator constrains the starting locations of its events to be uniformly distributed in the interior volume of the Marinelli beaker.
    The generator subsequently produces a direction with isotropic distribution and generates a primary vertex consisting of a single photon emitted from that point at the specified energy along the direction selected.
    This generator was used to simulate $2.5\times 10^7$ monoenergetic photons for each energy in the range $0.0002$ MeV to $12.0$ MeV in increments of $0.0002$ MeV.
    An example of the output for this generator can be seen in Fig. \ref{fig:en-resp-bkr}, demonstrating the response function for monoenergetic $1.173$ MeV photons (the lower of the two primary gamma rays emitted by $^{60}$Co).

\subsubsection{Constructing Isotope Responses from Energy Responses}
    In the cases where summing (due to either random or cascade coincidences) is negligible or minor (such as this work), response functions for isotopes can be constructed from individual monoenergetic photon response functions.
    The list of gamma-ray energies and per-decay intensities are extracted from the ENSDF for each isotope of interest.
    For each gamma ray, the simulated energy response function whose energy is closest to that of the gamma ray is multiplied by that gamma ray's intensity and summed into the isotope response function.
    An example of the per isotope energy deposition spectra generated by this technique is shown for $^{60}$Co in the blue histogram in Fig. \ref{fig:60Co-convolved}.

    The unfortunate feature of this technique is that it does not reproduce the summing peaks of random or cascade gamma-ray coincidences without significant additional work to calculate the appropriate scaling to produce the correct probabilities.
    However, while the summing peak is observed at low intensity in calibration spectra, it is of little importance relative to the primary peaks.
    The lack of summing intensity can be seen in the example detector calibration spectrum shown in Fig. \ref{fig:time-integrated-calibration-spectrum} where the strength of the $^{60}$Co summing peak is approximately $1$\% the strength of the higher energy $^{60}$Co line at $1.332$ MeV.

    Three groups of isotopes were chosen to be constructed for foreground components of the response matrix.
    The first group was the isotopes listed on the calibration source's certificate, not just those with significant remaining strength.
    These isotopes were $^{241}$Am, $^{109}$Cd, $^{57}$Co, $^{139}$Ce, $^{203}$Hg, $^{113}$Sn, $^{85}$Sr, $^{135}$Cs, $^{88}$Y, and $^{60}$Co.
    The second and third groups are relevant to later plans to analyze the effluent stream.
    The second group is the gaseous activation products.
    The isotopes in this group are $^{125, 125m, 127, 127m, 129m}$Xe and $^{41}$Ar.
    The third and final group are the gaseous fission daughters, namely radioisotopes of krypton, iodine, and xenon.
    These isotopes are $^{85, 87, 88, 89, 90}$Kr, $^{131,132, 133, 134, 135}$I, and $^{131m, 133, 133m, 135, 135m, 137, 138, 139}$Xe.

\subsection{Response Function Convolution and Response Matrix Construction}
\begin{figure}
    \includegraphics[width=\columnwidth]{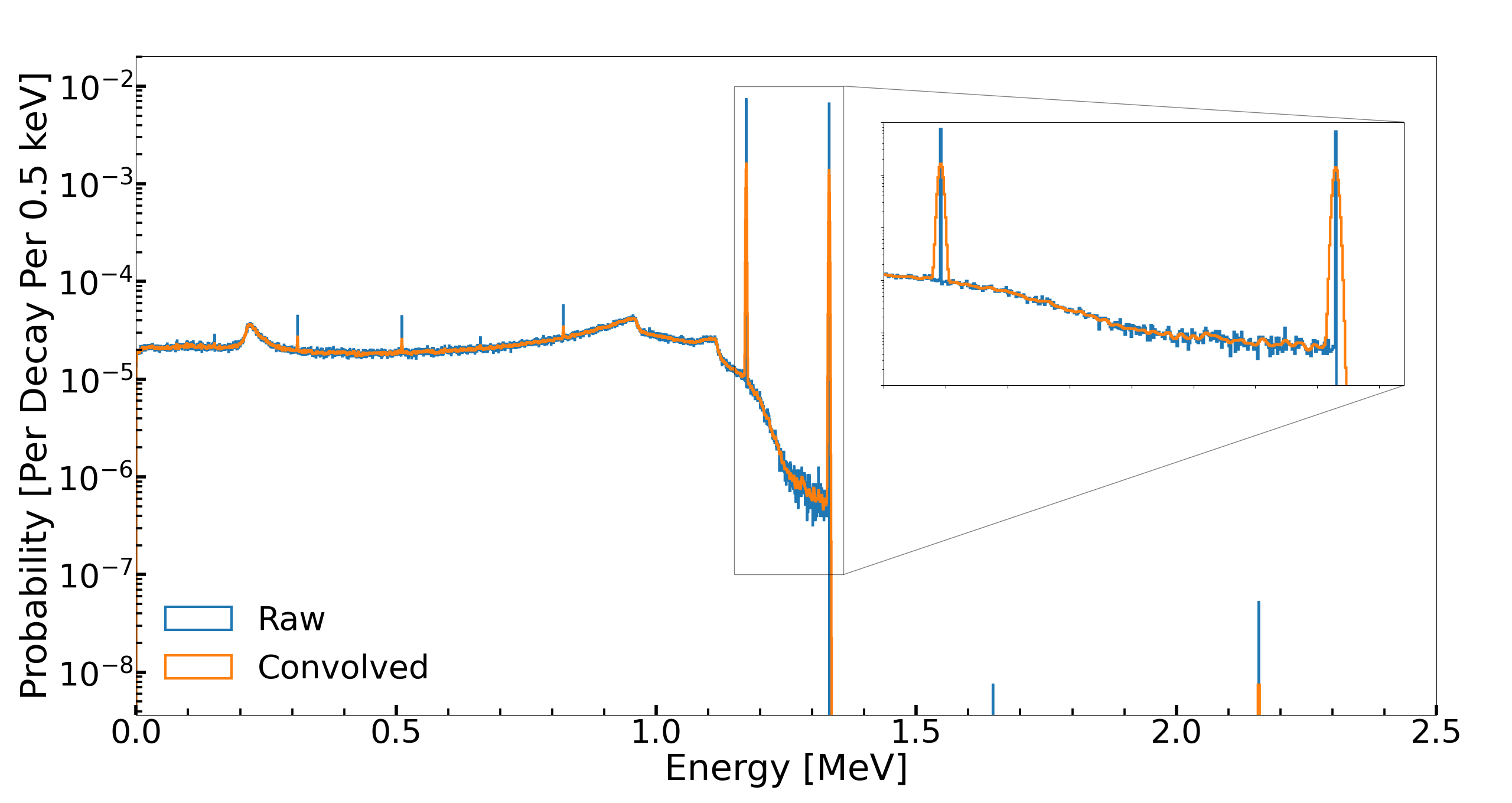}
    \caption{\emph{Color Online.} Raw and convoluted detector response for ENDSF constructed $^{60}$Co response.\label{fig:60Co-convolved}}
\end{figure}

    The spectra extracted directly from energy deposition simulations do not describe the detector response.
    This discrepancy is because they must be broadened to account for the finite resolution of the detector.
    Gaussian peak widths from energy calibration of the detector were fit with the empirically determined function:
\begin{align}
    \label{eqn:peak_std_dev_fit_func}
    \phantom{.}\sigma{}(E) = A + B \cdot{} \sqrt{E + C \cdot{} E^2}.
\end{align}
    Where $A$, $B$, and $C$ are the constants to fit, $E$ is the peak energy (during fitting, or the deposited energy during convolution,) and $\sigma(E)$ is the Gaussian width of that peak or bin.
    A Gaussian function whose width is set by \ref{eqn:peak_std_dev_fit_func} is then convoluted with each response function to yield the appropriately broadened response function.
    This convolution takes the form of:
\begin{align}
    \label{eqn:convolution_functions}
    \phantom{.}R_{conv}(E) = \sum_{\lambda{}=0}^{E_{max}} N(\lambda) \cdot e^{\frac{-(\lambda{}-E)^2}{2*\sigma{}(\lambda{})^2}} \cdot R_{sim}(\lambda{}).
\end{align}
    Where $\lambda$ is the energy bin in the deposited energy spectrum, $E$ is the energy bin in the convolved response spectrum, $N(\lambda)$ is the appropriate normalization to conserve probability between $R_{conv}$ and $R_{sim}$, and $\sigma{}(\lambda{})$ is the Gaussian width fit from \ref{eqn:peak_std_dev_fit_func}.
    An example of the input to (blue curve) and output from (orange curve) this procedure can be seen in Fig. \ref{fig:60Co-convolved}.

    Unfortunately, Eqn. \ref{eqn:convolution_functions} does not take the form of a conventional convolution which renders the efficient discrete convolution algorithm based on the fast Fourier transform, a discretized application of the convolution theorem\cite{convolutionTheorem}, useless.
    Instead to accelerate the convolution the summation is bounded. For each possible output energy $E$ the width function is examined to find the two values of $\lambda$ that are equivalent to $8\sigma{}$ in a conventional Gaussian by solving $\lambda{}-E = \pm{}8 * \sigma{}(\lambda{})$.
    These values of $\lambda$ are then used as bounds to the summation.
    The reduction in the summation size depends on the relative peak widths of the detector.
    For the HPGe detectors used in this work, this approximation reduced the number of terms to calculate by more than 95\%.

    After convolution, the response functions were inserted into a one-dimensional array of compound data structures in a hierarchical data format 5 (HDF5) file\cite{hdf5}.
    Each element of the array consisted of an identifier for the response function (proton number, mass number, decaying state energy), the response function itself, and the statistical errors of the response function, which have been propagated through the width convolution and combined with the errors in the width functions fit parameters $A$, $B$, and $C$.

    This method was employed on the foreground isotope response functions whose creation was described earlier.
    The background isotope response functions (discussed later) are handled differently.

\subsection{Other Empirical Corrections To Response Functions}
    Two additional empirical corrections to the response functions were necessary.
    The first issue was simple and wholly expected; at low energies (below ${\scriptstyle \sim}0.045$ MeV), the calibration of the detector and the trigger threshold for the electronics made the shape of the spectrum deviate from what is expected from the response functions.
    This low-energy departure was easily handled by simply truncating the spectra and response functions below ${\scriptstyle \sim}0.045$ MeV.

    The second, and more significant, issue was that the response functions were overestimating the low energy efficiency.
    This overestimation of low energy efficiency resulted in underestimating the number of decays that occurred for isotopes that primarily emitted low-energy gamma rays.
    To correct this issue, four calibration spectra (of the ${\scriptstyle \sim}68$ available) were selected to calculate an empirical correction.
    The dominant peaks were fit for each of these spectra to obtain their area, and then the peak efficiency was derived using the calibration source certificate.
    The relevant peaks in the response functions were similarly fit to obtain their peak efficiencies.
    The ratio of the simulated and actual efficiencies were then constructed for each peak and fit with the empirically determined function of:
    \begin{align}
        \label{eqn:emprical_eff_cor}
        \phantom{.}Ratio(En[MeV]) = 1 - \frac{1}{(En - a)^b}.
    \end{align}
    Here $a$ and $b$ are parameters of the fit, $En$ is the energy of the peak or the energy of the simulated monoenergetic photon in units of MeV.
    With $a$ and $b$ determined, each energy response function (and its error) was scaled by the ratio calculated with \ref{eqn:emprical_eff_cor} for its peak energy during the construction of the isotope response functions from the energy response functions, correcting the issue with overestimation of efficiency at low energies.

\subsection{Accounting for Background Terms in the Response Matrix}
\begin{figure}
    \includegraphics[width=\columnwidth]{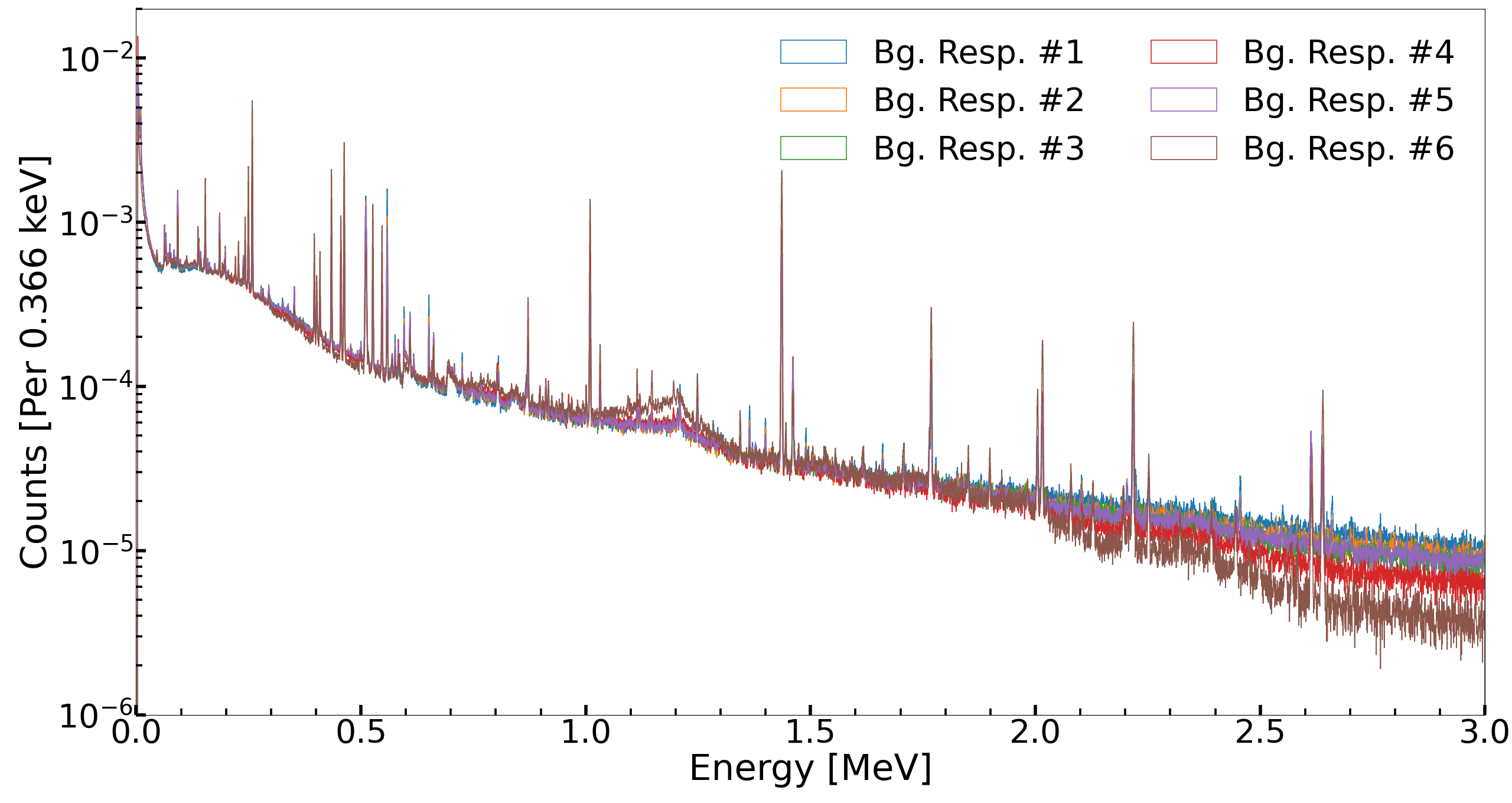}
    \caption{\emph{Color Online.} The six measured background response functions spread evenly across the time period of the data examined for this work. The differences at high energy are immediately obvious as are the shifts in the $^{40}$K Compton edge around $1.2$ MeV. Additionally, there are numerous subtle variations in relative peak intensities.  \label{fig:background_responses}}
\end{figure}

    Ideally, the response functions for background source terms are simulated and inserted into the response matrix.
    Handling background terms in this fashion has the distinct advantage of rendering the background components separable, handling shifts in their relative intensities.
    For instance, in the detector used for this work, the usual backgrounds from cosmic-ray muons, muon-spallation-induced fast neutrons, and naturally occurring radioactive material (NORM) are present.
    It is well known that the intensity and energy spectrum of cosmic-ray muons varies with the season and the solar cycle.
    Contrariwise, neglecting radon variations due to rain, NORM backgrounds for a particular location do not vary significantly with time.
    Having these components, fast neutron, cosmic-ray muon, and NORM separated is advantageous as then their intensity variations can be automatically accounted for in the decomposition process.

    Unfortunately, simulating the background, especially the neutron and cosmic-ray components, is significantly more computationally expensive than simulating the foreground due to several factors.
    First, many more primary particles need to be simulated as interactions with the active detector volume are rare.
    Also, neutrons are significantly more expensive to simulate than gamma rays, particularly once they thermalize.
    Finally, a much larger simulation volume is required for cosmic rays and the modifications of their flux by overburden (roofs, the stack, etc.)

    In the absence of precise geometric information, an additional cost of human time is spent on the simulation; because some components of the geometry, unimportant to the foreground spectra, would need to be iteratively refined to capture their effect on scattering and shielding correctly.
    For example, the effective shielding of the hole in the bottom of the shield by the detector electronics, the detector cold finger, and the liquid nitrogen dewar would take many trials to approximate.
    Such a refinement process costs numerous person-hours as well as CPU hours.

    The alternative used in this work is to capture several representative background samples and to place those directly in the response matrix.
    This method merely requires a significant period with no foreground signal in the detector.
    These spectra can be placed directly into the response matrix.
    As these are measured spectra, no convolution with width or corrections to efficiency is necessary.

    In this work, periods the stack was quiescent over the $\sim{}1.5$ years of data drawn from were identified, and six such background spectra were created.
    Each spectrum spanned approximately three months and was chosen out of the dataset when HFIR and REDC operations were not expected to produce effluent.
    A further cut was placed on the count rate of the spectrum, requiring that the count rate, averaged over a 10 second period, be no greater than the mean plus two standard deviations of the background count-rate of time known, by inspection, to have no effluent.
    The one-second spectra that passed these cuts were then processed in the same manner described later in \ref{sec:analysis-time-int}.
    Finally, to prevent the background weights from not being considered for decomposition convergence due to small weights, the spectra were normalized to have an integral of one, ensuring that the weights will be greater than one.
    These normalized background response functions, can be seen in Fig. \ref{fig:background_responses}.

    The cause of the elevation in the $1.2$ MeV region for ``Bg. Resp. \#6'' is currently unknown, but several scenarios are given that could account for the spectral variations.
    The first possibility is that the shield lid was not closed properly after an energy calibration.
    The second possibility is concurrent effluent emission from activities at the REDC that was low enough to be missed as a rate variation when evaluating the background data.
    The final possibility is that perhaps a sizeable LN2 dewar was moved for a time, temporarily slightly reducing the effective shielding around the detector.

%% file: analysis.tex
\subsection{Raw Data}
    Two acquisition systems are run in parallel from the dual outputs of the HPGe detector's preamplifier.
    ORNL uses the first system to acquire the data necessary for NESHAPS reporting.
    The data and analysis are not readily accessible because of the strict reporting requirements.
    However, access to the data and analysis during the weekly calibration procedure was allowed to compare the results obtained through the spectral decomposition presented in this work.
    The second acquisition provides data for monitoring different radionuclides outside the scope of the NESHAPS requirements for the HFIR and the REDC and provides the data for this research.

    The first acquisition system uses an Ortec DSPEC LF\cite{ortecDSPECLF} connected to the ``Energy'' output of the HPGe detector’s preamplifier.
    Ortec’s GammaVision\cite{ortecGammaVision} software is used to control the multi-channel analyzer in addition to data collection and analysis of stack effluent.
    As stated earlier, the detector is energy and efficiency calibrated with an Eckert \& Ziegler multi-nuclide ‘Simulated Gas in 130G GA-MA Gas Beaker’ standard providing photon energies ranging from 0$.0595$ to $1.3325$ MeV.
    This calibration is stored within GammaVision for analysis and verified weekly.
    GammaVision is programmed to continuously count the effluent while collecting and analyzing the background-subtracted spectrum every four hours.
    The analysis of the effluent spectra includes a library of radioactive noble gases and volatile nuclides that have been identified as potential emissions from processes at the REDC and HFIR.
    Results are submitted to the Clean Air Act Compliance Specialists in units of uCi/L of sample effluent for review of NESHAPS compliance.

    The second acquisition system takes the ``Timing'' output of the HPGe detector's preamplifier and routes it into an Ortec DSPEC Pro\cite{ortecDSPECPRO}.
    The DSPEC Pro digitizes the signal and acts as a multi-channel analyzer (MCA) to produce spectra.
    A custom data acquisition system, developed at ORNL, controls the amplifier and digitizer settings and the sample time of the DSPEC Pro.
    The energy spectra are stored every second to an SQLite\cite{sqlite2020hipp} database while simultaneously providing remote data backup and system health monitoring\cite{hpgeHfirEffluent}.

    The SQLite databases are read and processed by a custom Python\textsuperscript{\textregistered} \cite{PythonTR} script, which aggregates the one-second data into HDF5 files of four one hour length spectra.
    The detector is energy calibrated weekly, utilizing the photon lines of the radionuclides of the custom gas beaker source (see Table \ref{tbl:paraphrased-source-certificate}) and background photon peaks.
    The photon background peaks are one of the Pb x-rays at $0.085$ MeV, the annihilation photon – $0.511$ MeV, $^{40}$K – $1.46$ MeV, and $^{208}$Tl – $2.614$ MeV.
    The calibration photon peaks are fit using a Gaussian function and a constant plus a complementary error function to estimate the background continuum.
    A second-order polynomial function is used to perform energy calibration of the photon energy spectra converting the MCA channel number to energy.
    The calibration constants are saved to a file in NumPy format.
    When processing a new data file, the previous data file's energy calibration constants are loaded and applied to the current file.
    The current spectrum is analyzed using a simple integration technique to check if the calibration source is present.
    The spectrum is recalibrated if the source is present; otherwise, the previous energy calibration constants are used.

\subsection{Time Integration}
\label{sec:analysis-time-int}
\begin{figure}
    \includegraphics[width=\columnwidth]{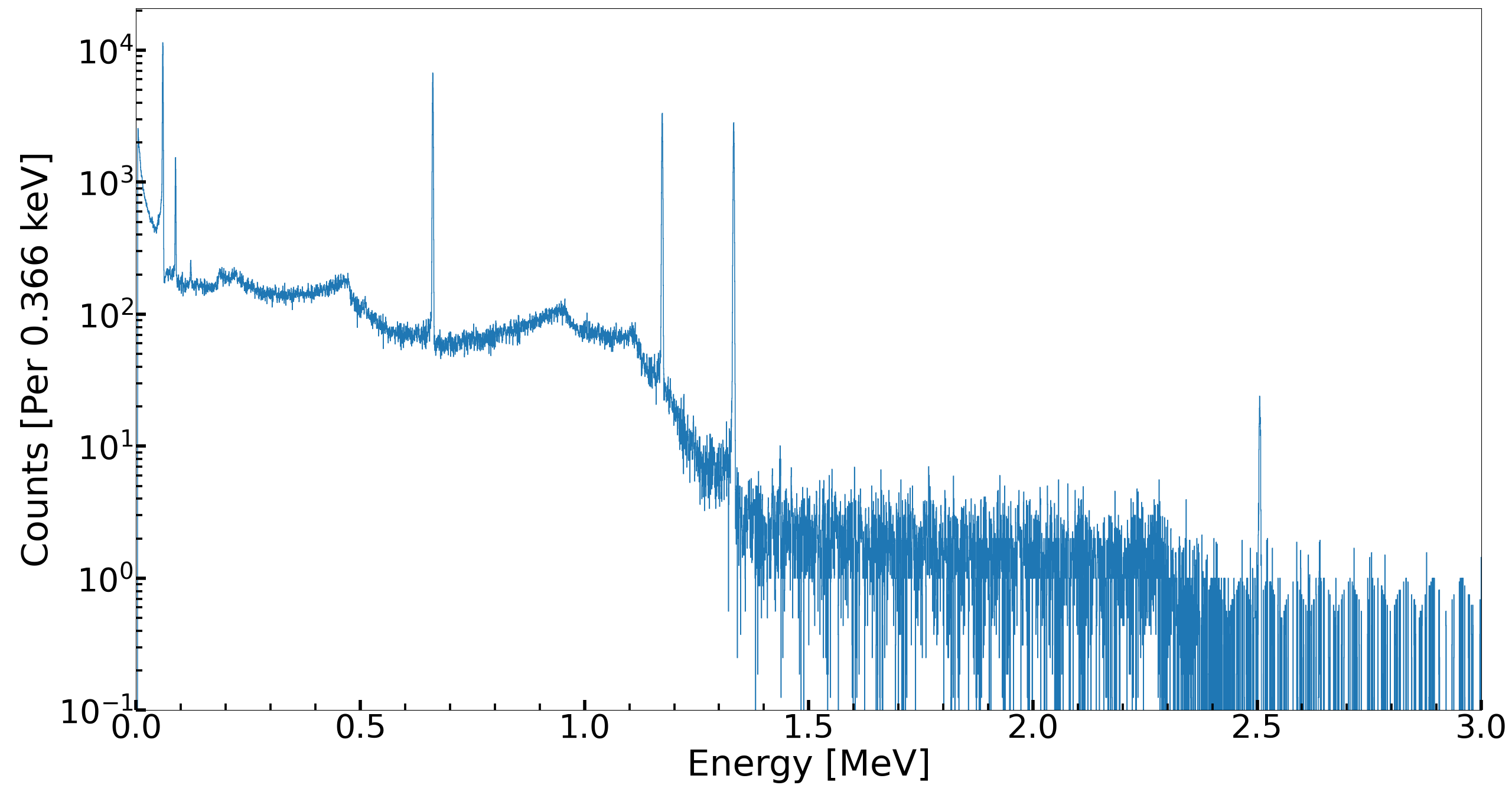}
    \caption{Time integrated calibration spectrum from 2021/02/19.
        Fractional counts occur because, during the time integration process, all spectra rebinned from its source calibration to be strictly linear with the low edge of the first bin at $0.0$ MeV and the high edge of the last be at $3.0$ MeV.
        \label{fig:time-integrated-calibration-spectrum}}
\end{figure}

\begin{figure}
    \includegraphics[width=\columnwidth]{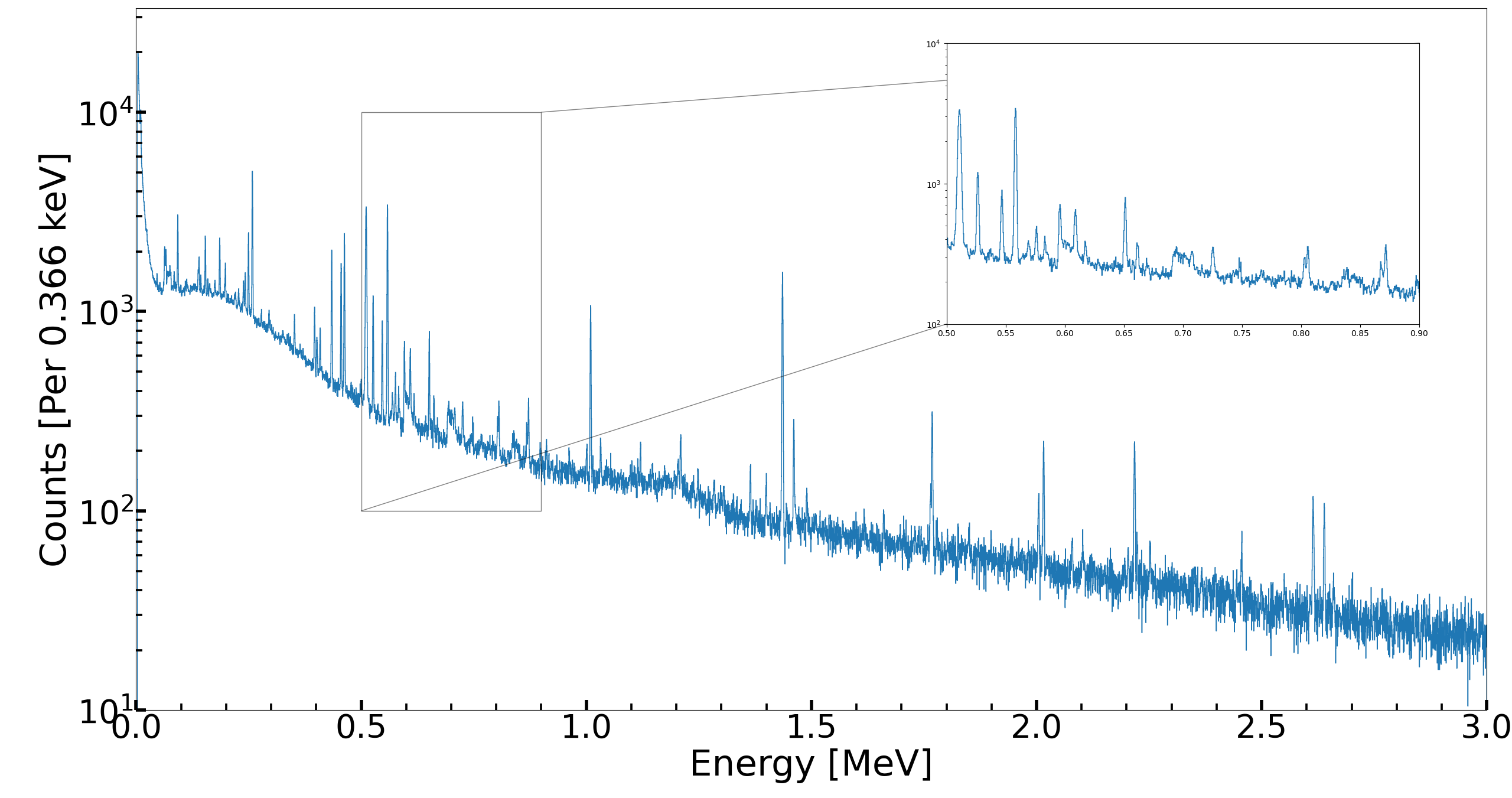}
    \caption{A background integral covering approximately a week of time in early October 2019.
        This spectrum was rebinned identically to Fig \ref{fig:time-integrated-calibration-spectrum}.
        The broad sawtooth-shaped peaks highlighted in the inset are due to neutrons (from cosmic-ray spallation) inelastically scattering from the germanium nuclei of the detector.
        The small plateau starting at ${\scriptsize \sim}0.563$ MeV is due to $^{76}$Ge$(n, n')$ exciting the first $2^+$ state of $^{76}$Ge.
        The sawtooth at $0.595$ MeV is from $^{74}$Ge$(n, n')$ exciting the first $2^+$ state of $^{74}$Ge.
        The sawtooth at $0.691$ MeV is from $^{72}$Ge$(n, n')$ exciting the first non-ground $0^+$ state of $^{72}$Ge.
        The final, small, sawtooth starting at $0.834$ MeV is also from $^{72}$Ge$(n, n')$ exciting the first $2^+$ state of $^{72}$Ge.
        \label{fig:time-integrated-background-spectrum}}
\end{figure}

    As stated previously, the data obtained directly from the system are energy spectra for every second.
    For each second, the uncalibrated energy spectrum, the live time for that spectrum, and the timestamp are available.
    Integration is necessary to build spectra with sufficient statistics for decomposition time.
    The integration is accomplished by finding boundary timestamps for the calibration data sets using the count rate, and the one-second spectra within those boundaries are processed as described below.

    First, the spectrum's bins are calibrated with the parameters stored in the file.
    Next, the spectrum and its errors are live time corrected according to the spectrum's live and real times.
    Then, the spectrum is rebinned into the spectral binning that matches the response functions ($8192$ bins with the lowest bin edge as $0.0$ MeV and the highest bin edge as $3.0$ MeV).
    The rebinning method is a variation of the method given in \cite{knollRadDetAndMeasurement} for reassigning counts in a spectrum.
    Instead of using a polynomial of degree $8191$, a piece-wise stepped function is defined whose value between bin edges of the original histogram is a constant fixed to the number of counts in the bin.
    This choice of function has the benefits of preserving the total number of counts, preserving the average energy of the spectrum, and converging with the slower stochastic rebinning methods in the high statistics limit.
    Finally, the rebinned spectrum is added to the integral spectrum.
    An example spectrum generated this way for a source calibration run can be seen in Fig. \ref{fig:time-integrated-calibration-spectrum}.
    A, compared to the spectra used in the response matrix, short integration of background data, when the stack was quiescent yielded the spectrum seen in Fig. \ref{fig:time-integrated-background-spectrum}.

    After each calibration period was integrated, it was added to a one-dimensional array of compound data structures in an HDF5 file.
    Each array element contains the start, end, and mid times for the spectrum, the time-integrated spectrum, and the spectrum's errors.

    \subsection{Decomposition}
    A batch decomposition program was written in C++ to decompose the time-integrated spectra using the convolved and corrected response matrix.
    The decomposition algorithm itself was implemented using version 3.3.9 of the Eigen library\cite{eigenweb}.
    The complete propagation of errors from the time-integrated spectrum \emph{and} the response matrix (as described in the Appendix) was also implemented, and errors are propagated for each decomposed spectrum after convergence has been obtained using the penultimate iteration's weights in the calculation.
    Since the decomposition of a spectrum has no dependence on any other decomposition, the program was written to use multiple threads of execution to take advantage of the embarrassingly parallel nature of the task.

    The program takes a configuration file as input, allowing the user to specify the input response matrix and data file.
    Additionally, the file allows specification of where input spectra and response functions should be truncated prior to decomposition (handling the truncation necessary to accommodate low energy threshold and calibration effects).
    Finally, the configuration file also specifies the convergence criterion mentioned in \ref{subsec:inverse-problem}.
    For this work, the low energy truncation of the spectrum was set to be everything below $0.045$ keV, the consideration threshold was set to $10^{-6}$, and the maximum relative change for convergence was set to $0.005$.

    For every spectrum it decomposes, the program outputs a compound data type in an array written using the HDF5.
    Each array element contains the vector of weights obtained for the corresponding input spectrum and the covariance matrix for those weights.
    For convenience, the batch decomposer also places the original response matrix and input spectra in the output HDF5 file.

%% file: results.tex
\subsection{Decomposition of Calibration}
\begin{figure}
    \includegraphics[width=\columnwidth]{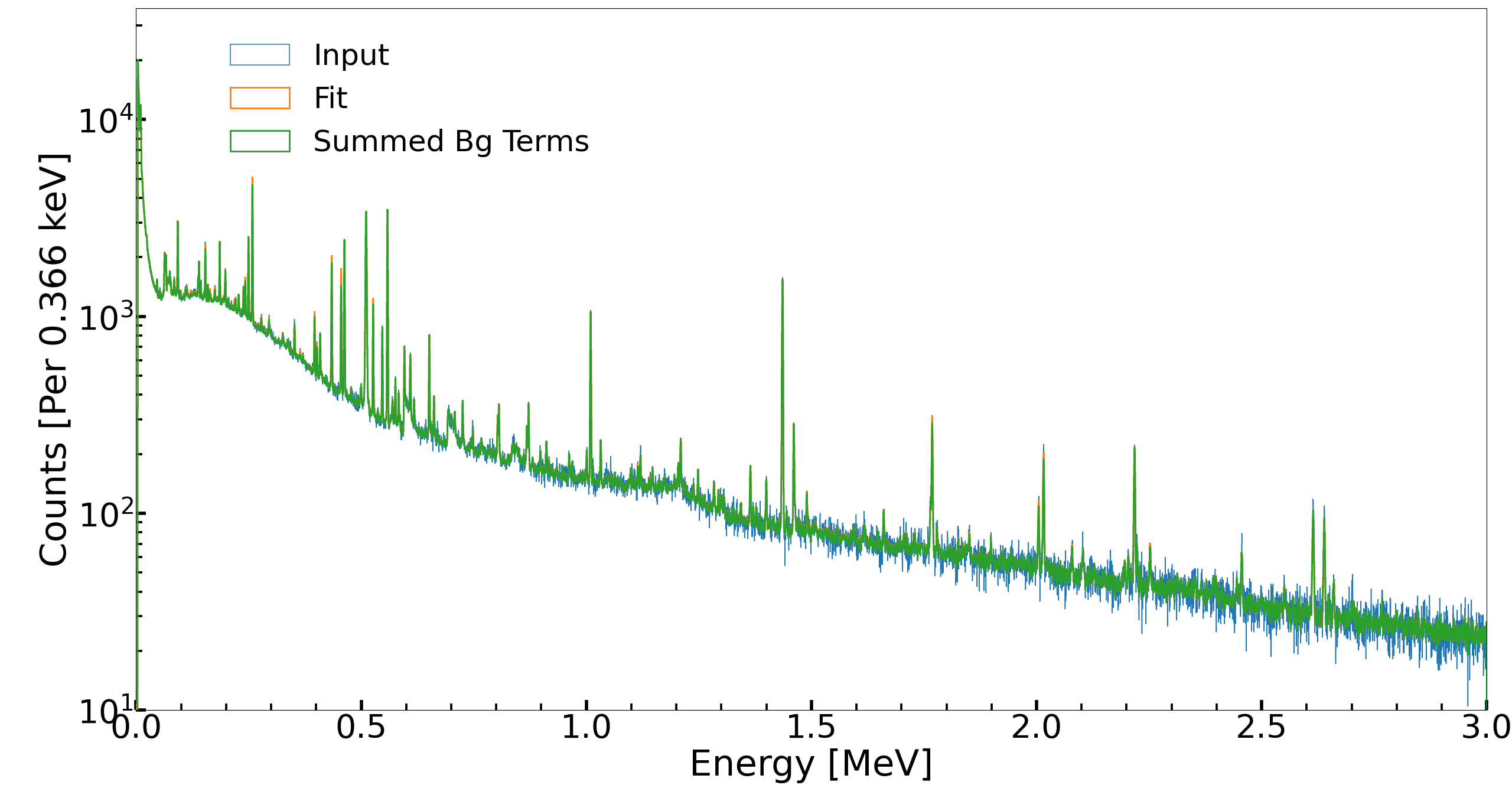}
    \caption{\emph{Color Online.} Decomposition of the background spectrum shown in Fig. \ref{fig:time-integrated-background-spectrum}.
        The input spectrum is blue, and the sum of all the response functions scaled by their weights is orange.
        The green curve is the sum of the scaled background terms.
        The degree of overlap between the background and sum curves shows that the decomposition weights the background functions very highly and assigns very little to the foreground responses.
        \label{fig:bg_decomp_example}}
\end{figure}

\begin{figure}
    \includegraphics[width=\columnwidth]{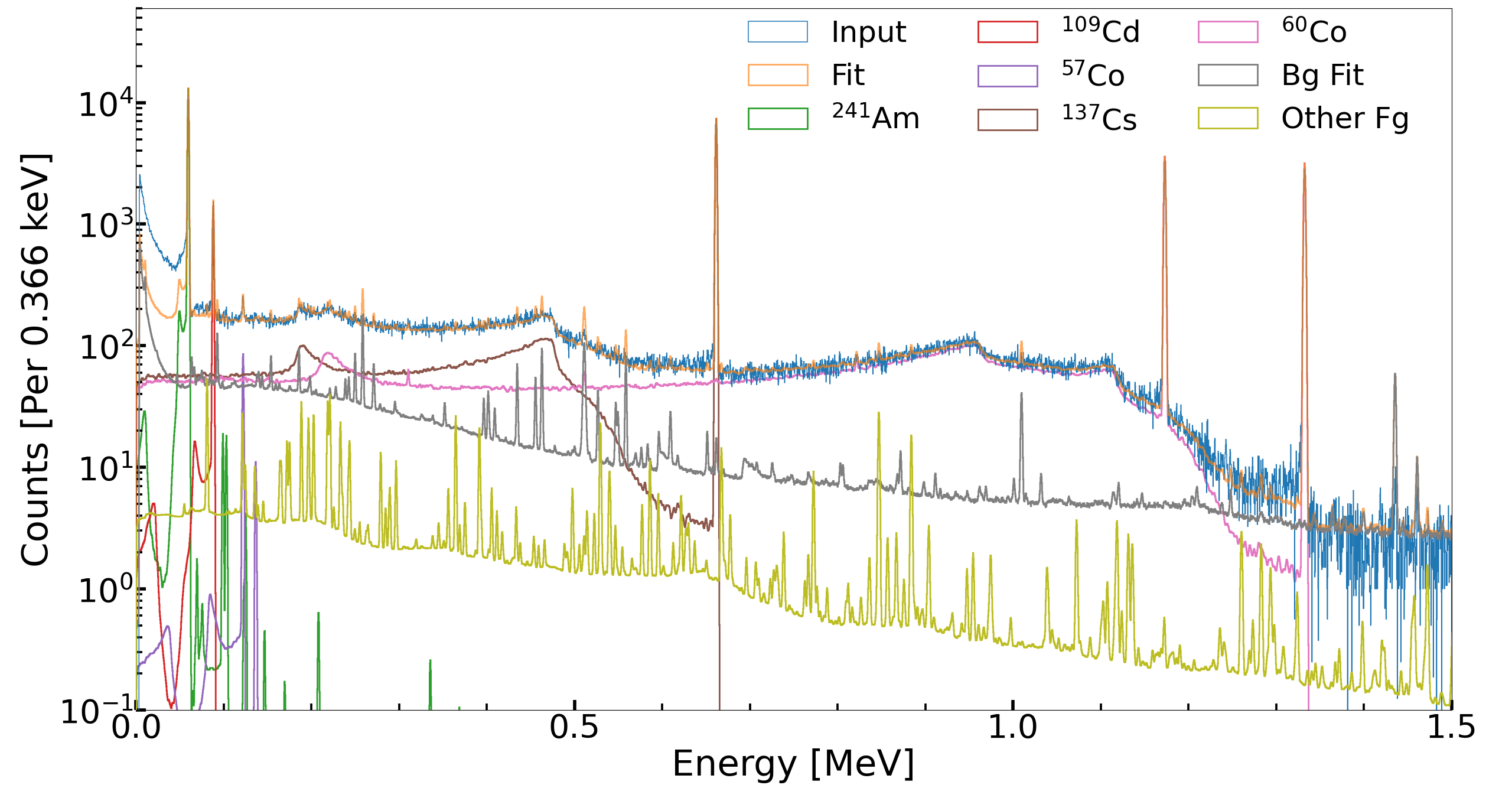}
    \caption{\emph{Color Online.} Decomposition of the calibration data shown in Fig. \ref{fig:time-integrated-calibration-spectrum}.
        The input spectrum is blue, and the sum of all the response functions scaled by their weights is orange.
        The gray curve is the sum of the scaled background terms.
        The dark yellow curve is the sum of all foreground components that are not the remaining isotopes in the calibration source.
        Finally, the scaled response functions for the remaining components of the calibration source are also shown. \label{fig:decomp_example}}
\end{figure}

    Calibration spectra from mid-June 2019 to mid-February 2021 and a summed background drawn from the same period were processed to test the decomposition method.
    Sometimes, despite the beaker not being in place, there was a noticeable effluent signal.
    In other cases, the calibration was too short, and the $^{57}$Co peak was not visible or was subject to significant statistical error.
    In a few cases, the detector was experiencing significant noise, putting the count rate outside the range expected for calibration.
    In these cases, the calibration spectra were discarded from this analysis, leaving $68$ spectra to decompose.
    Approximately a week of background signal was summed and decomposed to test the effectiveness of the background functions.
    Fig. \ref{fig:bg_decomp_example} shows the decomposition of the background.
    Fig. \ref{fig:decomp_example} shows an example of the fit to calibration data yielded by the decomposition.
    In both cases, the sum line is calculated by scaling response functions by their weights and summing them.
    The background line is calculated by only summing the weighted empirical background terms of the response matrix.
    The most significant deviation between the sum and input is in the lower energy region (${\scriptstyle \sim}0.045$ MeV and below).
    The deviation is likely due to increases in background components because the lid must be open during calibration (to allow the hoses and Marinelli beaker to hang outside the shield to not interfere with the calibration.)
    The weights represent the number of isotope decays represented by their response functions.

    With this work's convergence parameters, response matrix, and spectra, the decomposition process required $1715 \pm 432$ iterations on average.
    Using an AMD Ryzen\textsuperscript{\texttrademark} Threadripper\textsuperscript{\texttrademark} PRO 3995WX workstation and the linux command-line utility \verb*|time| average times for the full decomposition, error bar calculation with only spectrum errors, and the full error calculation were extracted.
    Decomposition without error-bar calculation requires, on average, $160$ ms per input spectrum.
    Calculation of errors using only the input spectrum errors requires approximately $23$ ms per input spectrum.
    Calculation of errors using input spectrum errors and response matrix errors requires $\sim{}1851$ ms per input spectrum.
    Usage of the Linux profiling tool \verb*|perf| suggests that cache misses are the primary reason for the slowness of the full error calculation.
    A careful examination of the Eigen library's sparse matrix methods may yield methods to improve the calculation's cache coherency, giving a significant speedup.

    Inspection of Fig. \ref{fig:bg_decomp_example} shows that the decomposition process passes a critical sanity check; namely, the empirical background response functions allow nearly perfect decomposition of pure background data.
    In the case of a few background peaks, the fit line emerges slightly above the background line showing that there were minor relative intensity shifts between the short background integration and the long ones used to generate the response functions.
    While this issue could potentially be ameliorated by using more background response functions with shorter integration times, this would cause the responses to be less smooth at high energy, causing other issues.

    Upon examination of Fig. \ref{fig:decomp_example}, a few small peaks appear in the fit spectrum that are either absent or only weakly present in the input spectrum, namely $0.258$, $0.463$, $0.511$, $0.558$, and $1.436$ MeV.
    Despite this, the agreement between the input and the fit is excellent.
    The small peaks appear to be almost exclusively from empirical background peaks being more intense than in the calibration spectrum's background.
    The discrepancy is likely due to some combination of the following three factors.
    The first possibility is that there may well be tiny amounts of particulate decay daughters of gaseous effluents on the inside of the Marinelli beaker.
    If this is the case, perhaps a clean Marinelli beaker could be obtained, and background spectra obtained using it.
    The second possibility is a background shift because the lid of the lead shield must be kept slightly open during calibration to accommodate the hoses of the Marinelli beaker.
    The third possibility is that the calibration source contributes a small amount more shielding of the background than the Marinelli beaker.
    However, given the source's very low density and the energy of some of the peaks present, this explanation seems unlikely.

    On further inspection of Fig. \ref{fig:decomp_example} shows all other foreground components are strongly suppressed relative to the calibration foreground and even the background.
    The sum of all of these components is approximately an order of magnitude below the background.
    Further, this sum is at least that much below the calibration nuclide response functions, frequently more.
    In a few cases, a particularly intense peak from these foreground isotopes gives a tiny peak in the sum spectrum, but these are minor.

\begin{figure}
    \includegraphics[width=\columnwidth]{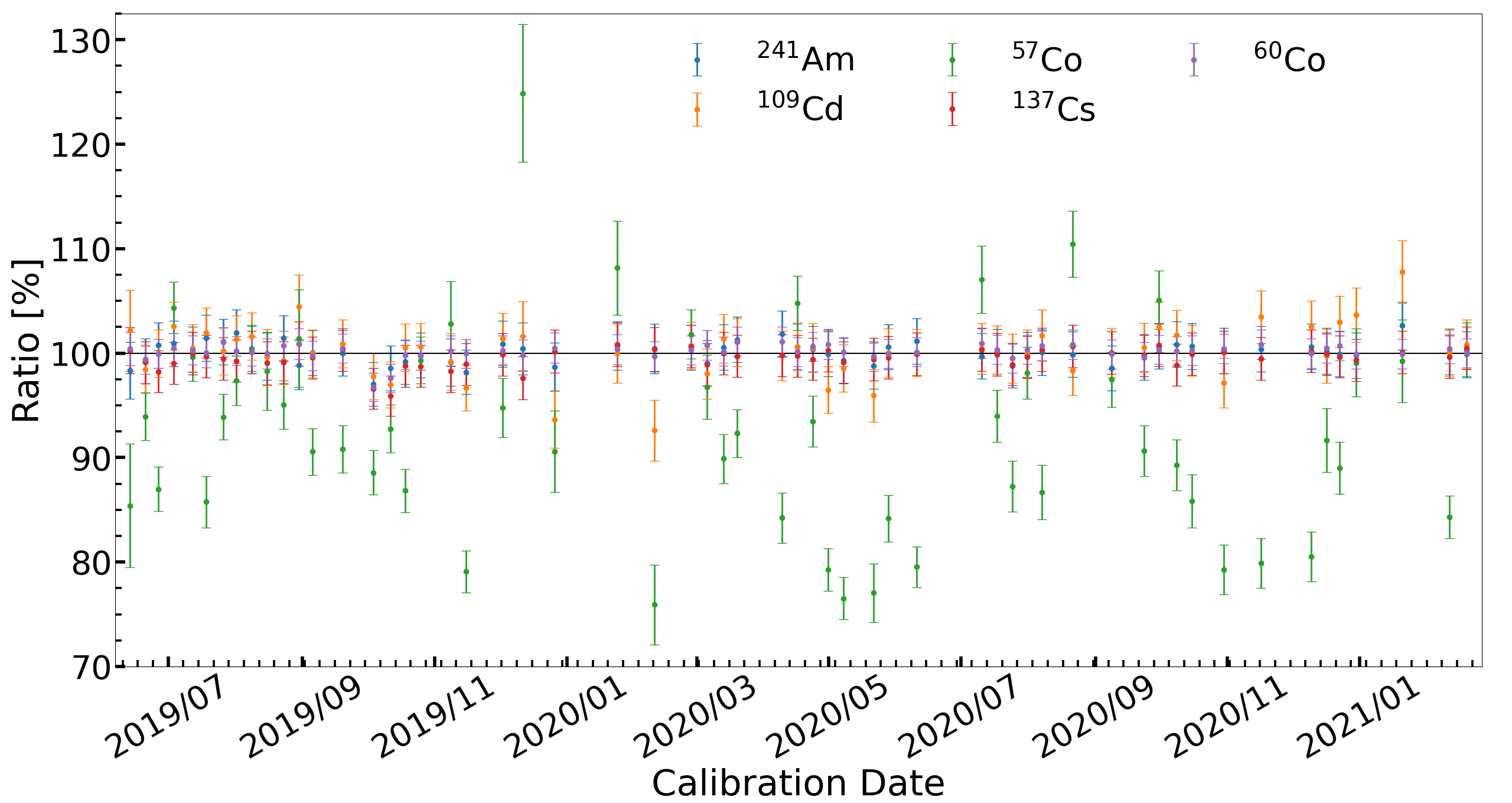}
    \caption{\emph{Color Online.} The ratio presented as a percent of the number of decays found by the decomposition process to the calculated expected decays from the source calibration certificate.
        The gaps in the data are due to several factors explained in the text, including low count statistics, excessive detector noise, etc.
        Despite these gaps, the ratios fall in a narrow band close to $100$\%,excepting for $^{57}$Co, showcasing the stability and accuracy of the decomposition technique.\label{fig:decomp_ratios}}
\end{figure}

    \begin{table*}
        \caption{\label{tbl:calibration-decomp-results}
            A table for several calibration runs of average activities derived in 3 ways; decomposition, source certificate calculation, and peak fitting from Ortec's GammaVision software. The dates have been chosen randomly to span the time the analysis was performed while being approximately equidistant and sorted in ascending order.
            All errors provided are $1\sigma$.
            The ``Decomp.'' columns are the decomposition calculated response function weight (and error) divided by the calibration duration.
            The ``Calc.'' columns are the decay corrected activities in the source certificate divided by the calibration duration, providing an average activity.
            The ``Report'' columns are the obtained from the GammaVision analysis contained in the calibration reports.
            Please note that the units of activity for a given row's entries are given in the start of the row (a consequence of the broadly varying source intensities).}
        \begin{center}
            \begin{tabular}{c||c|c|c||c|c|c||c|c|c}
                \hline
                & \multicolumn{3}{c||}{2019-06-13}         & \multicolumn{3}{c||}{2019-10-11}         & \multicolumn{3}{c}{2020-02-10}        \\
                \cline{2-10}
                                 & Decomp.    & Calc.      & Report         & Decomp.    & Calc.      & Report         & Decomp.    & Calc.      & Report       \\
                \hline
                $^{241}$Am [kBq] & $1.92 (4)$ & $1.96 (4)$ & $2.2  (1)$     & $1.93 (1)$ & $1.96 (4)$ & $2.16 (10)$    & $1.96 (2)$ & $1.95 (4)$ & $2.23 (10)$  \\
                \hline
                $^{109}$Cd [kBq] & $2.34 (7)$ & $2.29 (5)$ & $2.39 (165)$   & $1.86 (2)$ & $1.91 (4)$ & $1.93 (164)$   & $1.48 (4)$ & $1.59 (3)$ & $1.61 (165)$ \\
                \hline
                $^{57}$Co [Bq]   & $8.2  (6)$ & $9.7  (2)$ & $9.9  (524)$   & $6.6  (1)$ & $7.1  (1)$ & $7.8 (688)$    & $4.0  (2)$ & $5.2 (1)$  & $5.4 (798) $ \\
                \hline
                $^{137}$Cs [Bq]  & $685  (7)$ & $693 (14)$ & $695  (21)$    & $659  (2)$ & $687 (14)$ & $666  (20)$    & $685  (4)$ & $683 (14)$ & $691 (21) $  \\
                \hline
                $^{60}$Co [Bq]   & $693  (5)$ & $690 (10)$ & $668  (29)$    & $645  (2)$ & $661  (9)$ & $625  (28)$    & $630  (3)$ & $632  (9)$ & $613 (29) $  \\
                \hline
                \hline
                & \multicolumn{3}{c||}{2020-06-10}         & \multicolumn{3}{c||}{2020-10-15}         & \multicolumn{3}{c}{2021-02-19}          \\
                \cline{2-10}
                                 & Decomp.    & Calc.      & Report         & Decomp.    & Calc.      & Report         & Decomp.    & Calc.      & Report         \\
                \hline
                $^{241}$Am [kBq] & $1.98 (7)$ & $1.95 (4)$ & $2.20  (10)$   & $1.97 (1)$ & $1.95 (4)$ & $2.19 (10)$    & $1.95 (1)$ & $1.95 (4)$ & $2.20 (10)$    \\
                \hline
                $^{109}$Cd [kBq] & $1.33 (1)$ & $1.33 (3)$ & $1.39 (177)$   & $1.10 (2)$ & $1.10 (2)$ & $1.15 (178)$   & $0.92 (1)$ & $0.91 (2)$ & $0.99 (196)$   \\
                \hline
                $^{57}$Co [Bq]   & $3.04 (6)$ & $3.83 (7)$ & $3.47 (107)$   & $2.38 (6)$ & $2.77 (6)$ & $2.87 (9988)$  & $2.01 (4)$ & $2.00 (4)$ & $2.59 (21848)$ \\
                \hline
                $^{137}$Cs [Bq]  & $677  (2)$ & $677 (14)$ & $685  (21)$    & $671  (2)$ & $672 (14)$ & $674  (21)$    & $669  (2)$ & $667 (13)$ & $669 (21)$     \\
                \hline
                $^{60}$Co [Bq]   & $606  (1)$ & $605  (8)$ & $588  (29)$    & $579  (2)$ & $578  (8)$ & $563  (29)$    & $552  (1)$ & $552  (8)$ & $539 (30)$     \\
                \hline
            \end{tabular}
        \end{center}
    \end{table*}

    For comparison against ground truth, the number of decays was calculated from the source certificate.
    The decay corrected activity of each isotope on the source certificate was integrated between the start and stop times of the spectrum.
    The ratio (expressed as a percent) of the isotope response function's weight found through decomposition to the calculated decays per isotope from the source certificate per calibration period is shown in Fig. \ref{fig:decomp_ratios}.
    The variability of the $^{57}$Co results is easily explainable as it is the least intense source that remains measurable in the calibration standard.
    The second weakest source (as of the start of the calibration spectra decomposed here,) $^{60}$Co, was ${\scriptstyle \sim}71$ times stronger than $^{57}$Co at the start of this series of measurements and was ${\scriptstyle \sim}280$ times stronger at the end of the series.

    The number of decays in the period was divided by the duration in seconds to obtain the average activity of each radionuclide in the calibration standard.
    Due to the short time duration of the calibration procedure relative to half-lives of the isotopes, the average activity should be approximately equal to the activity of the isotope at the start (or end) of calibration.
    Therefore, activity extraction from the isotope weights does not require decay correction.
    Average activities from the calibration source certificate were obtained by integrating the decay corrected activities across the appropriate period and dividing that by the length of the period.

    Since the calibration reports generated by GammaVision in the NESHAPS reporting data acquisition system were made available for this study, they were compiled to contain the same calibration periods as the decomposition analysis.
    Table \ref{tbl:calibration-decomp-results} contains six randomly selected but approximately equidistant dates that span the overall time period the decomposition analysis was performed over (i.e., the x-axis in Fig. \ref{fig:decomp_ratios}).
    The table provides a means to compare the absolute radionuclide activities found from the decomposition method (``Decomp.''), the calculated activity from the source certificate (``Calc.''), and the activities from the Gammavision software reports (``Report'').
    Generally, the decomposition results are stable, have small errors, and agree with the calculated values at $1\sigma$.
    There are a few outliers, particularly comparing the results for $^{57}$Co.
    What is very apparent are the significant errors associated with the report values found using GammaVision.
    Since that analysis is a traditional peak fitting technique, it is evident that the error associated with the radionuclide activity is highly dependent on the counting statistics (i.e., count time). Some of the reported error is so elevated that the results would be discarded.
    However, since it is not solely dependent on peak counting statistics, the decomposition method produces reasonable errors and preserves the integrity of all calibration isotopes used in the calibration standard.

\subsection{Decomposition Sensitivity}

    \begin{figure}
        \centering
        \includegraphics[width=\columnwidth]{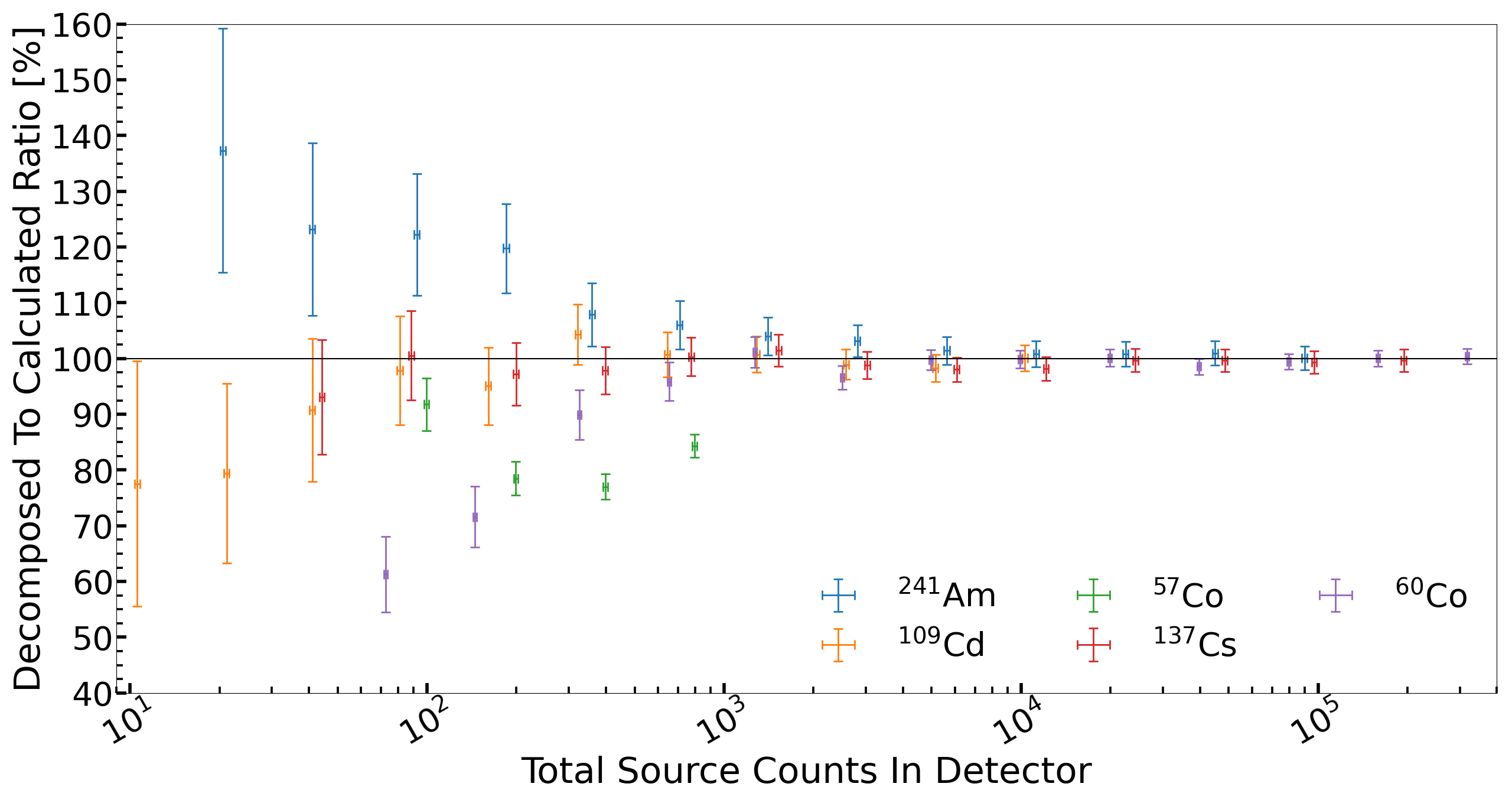}
        \caption{\emph{Color Online.} The ratio expressed as a percent of the total decays extracted by the decomposition process to the decays expected by calculation from the source calibration certificate as a function of the number of counts of each radionuclide.}
        \label{fig:sensitivity_study}
    \end{figure}

    The longest calibration measurement was subdivided in time and decomposed to test the sensitivity of the decomposition method.
    That is to say that the same calibration was integrated over progressively smaller periods, and integrated spectra, each with fewer statistics than the last, were decomposed.
    The decompositions of the subdivided spectra were then used to calculate the ratio of the decomposition weight to the calculated decays.
    For each nuclide, the expected number of decays in that period was multiplied by the detection probability per decay (the sum of the response function for that nuclide).
    This calculation produces an approximate number of counts attributable to that source and is plotted as the ``Total Source Counts In Detector'' for that source in each spectrum, and in Fig. \ref{fig:sensitivity_study} the ratio versus that value is plotted.

    As can be seen qualitatively in Fig. \ref{fig:sensitivity_study} the method produces excellent results when ${\scriptstyle \sim}1000$ or more counts from a source are present in the spectrum.
    Further, at and above ${\scriptstyle \sim}250$ counts from a source in the spectrum results are within $10$ \% - $25$ \% of the calculated activity.
    Given that the $^{57}$Co points for all the calibrations lie between ${\scriptstyle \sim}90$ and ${\scriptstyle \sim}1000$ counts, this explains the significant variability seen in Fig. \ref{fig:decomp_ratios}.

    These results suggest a method of ensuring the time-integrated spectrum acquired during monitoring is long enough.
    Using the decomposed weight of all isotopes whose weights are above some threshold, one can determine if the spectrum has sufficient statistics for that weight; if not, the spectrum's integration time can be extended until it crosses the threshold.
    Alternately one could use the sensitivity study to determine which weights to mark as suspect.

%% file: conclusion.tex
    Concluding, a decomposition technique was developed and applied to high-resolution gamma-ray spectra.
    A high-fidelity model of an HPGe detector was developed.
    Energy deposition in the model due to foreground photon sources were simulated from 0.0002 to 12.0 MeV in 0.0002 MeV steps.
    These simulations built an inventory of simulated photon energy response functions.
    With this inventory, the response due to radionuclides could be calculated using the nuclear decay data.
    Then these foreground response functions and empirically determined background response functions were used in combination with the MLEM method to decompose spectra.
    Finally, the MLEM was shown to be a good method for isotope identification and quantification.

    In general, the decomposition weights were shown to be in good agreement with the source certificate.
    The decreased accuracy of the $^{57}$Co results can be attributed to the age of the calibration source and the reduced radioactivity of the isotope.
    In more than $70$\% of cases, the decomposition calculated activities are closer or no further from the certificate value than GammaVision.
    Further, with poor peak statistics, the error of the MLEM value does not suffer as dramatically as GammaVision's, which can be seen in the more than $20$\% of cases where GammaVision value was closer to the certificate activity, but the error bar was so large it rendered the value useless.

    In future applications of this work, several improvements can be made.
    First, geometry relevant to the background terms should be determined, and the various components of the background (cosmic-ray muons, cosmic-ray spallation neutrons, and NORM) should be simulated.
    Second, those background components should be simulated with the shield's lid open and closed to replicate usage scenarios better.
    Finally, the detector geometry should be tweaked to bring the simulated efficiencies closer to those measured.
    Doing this may reduce the need for \emph{post-facto} corrections.
    A closer initial match may allow greater decomposition accuracy even if \emph{post-facto} corrections are still necessary.

%% file: appendix_a.tex
\subsection{Uncertainties in Spectrum}
    In the simple case where errors are only present for the observed spectrum $\vec{O}$, the errors in $\vec{I}$ can be determined from the final iteration of the maximum likelihood decomposition as follows.
    First by observing that for a given iteration \ref{eqn:decomp_eqn} can be expressed as a simple vector equation:
\begin{align}
    \label{eqn:simplified_err_start_eqn}
    \vec{I}^{s+1} = \mathbf{M^{s}} \vec{O}
\end{align}
    where $\vec{I}^{s+1}$ is the vector of weights from the final iteration, $\vec{O}$ is the observed spectrum, and $\mathbf{M^{s}}$ is the intermediate matrix defined as:

\begin{align}
    \label{eqn:simplified_err_M_eqn}
    M^{s}_{\mu{}i} = \frac{1}{\sum_{j}R_{\mu{}j}}\frac{I_{\mu{}}^{s}R_{\mu{}i}}{\sum_{\alpha{}}R_{\alpha{}i}I_{\alpha}^{s}}
\end{align}
    Here $\mu$ and $\alpha$ are row indices for the response matrix / weight vectors, $j$ and $i$ are column indices for the response matrix, $R$ is the response matrix, and $I^{s}$ is the vector of weights from the penultimate iteration of the decomposition.

    Then by noting that the covariance matrix for a linear equation such as \ref{eqn:simplified_err_start_eqn} can be calculated using:
\begin{align}
    \label{eqn:simplified_covariance_calc}
    \mathbf{\Sigma^{I^{s+1}}} = \mathbf{M^{s}} \mathbf{\Sigma^{O}} \mathbf{M^{s}}^{T}
\end{align}
    Here $\mathbf{\Sigma^{I^{s+1}}}$ is the covariance matrix of the calculated input coefficients $\vec{I}^{s+1}$ and $\mathbf{\Sigma^{O}}$ is the covariance matrix of the observed spectrum $\vec{O}$ (which, assuming the bins of the observed spectrum are uncorrelated is a diagonal matrix of the squared errors in each spectrum bin of $\vec{O}$).
    To calculate the final covariance matrix of  $\vec{I}^{s+1}$ one merely needs to calculate $\mathbf{M^{s}}$ using the next-to-last iteration's parameters and plug it in to \ref{eqn:simplified_covariance_calc}.
    Once the covariance matrix has been obtained, the uncorrelated errors in the weights are found by taking the square roots of the diagonal coefficients.
    Alternatively, the entire covariance matrix may be retained for further error propagation.

\subsection{Uncertainties in Spectrum \emph{and} Response}
    Error propagation in the situation where errors in the response matrix are also known or estimated is more complex than otherwise; but still, in theory, calculable.
    Previously, with errors only in the spectrum, we relied on the fact that the Jacobian matrix of the equation was identical to the intermediate matrix.
    Here the Jacobian matrix is significantly larger, if the response matrix is $n\times{}m$ then the Jacobian matrix for both errors in the response and the input spectrum is $n\times{}(n+1)\cdot{}m$.
    Thus, in practice, the Jacobian matrix grows accordingly for large response matrices.
    This excessive size can be ameliorated somewhat by storing the Jacobian sparsely; for the spectra and matrices used in this work, the Jacobians were ${\scriptstyle \sim}34$\% non-zero.

    To calculate the covariance matrix, one starts with the fact that, to first order, error propagation through any vector equation $\vec{f}(\vec{x})$ can be written as:

\begin{equation}
    \label{eqn:err_with_jacobian_eqn}
    \mathbf{\Sigma}^{\vec{f}} = \mathbf{J} \mathbf{\Sigma}^{\vec{x}} \mathbf{J}^{T}
\end{equation}
    where $\mathbf{\Sigma}^{\vec{f}}$ is the covariance matrix of $\vec{f}$, $\mathbf{\Sigma}^{\vec{x}}$ is the covariance matrix of the input parameters, and $\mathbf{J}$ is the Jacobian matrix of $\vec{f}$ and the Jacobian matrix is defined to be: $J_{ij} = \frac{\partial f_i}{\partial x_j}$ (where $J_{ij}$ is the coefficient of the $i$th row and $j$th column of the Jacobian matrix, $f_i$ is the $i$th element of $\vec{f}$ and $x_j$ is the $j$th argument to $\vec{f}$).
    Applying the definition of the Jacobian matrix to Eqn. \ref{eqn:decomp_eqn} requires some effort but can be done as follows.
    First the ordering of arguments to the decomposition function must be defined, in this work the convention chosen is as follows:

\begin{equation}
    \label{eqn:variable_naming}
    x_j = \begin{cases}
        R_{\lfloor \frac{j}{m} \rfloor ,j \% m} & j < p\\
        O_{j-p} & p \le{} j < p + m
    \end{cases}
\end{equation}
    Where $p = m \times{} n$, $m$ is the number of bins in each response function, $n$ is the number of response functions, $R_{\lfloor{}\frac{j}{m}\rfloor{},j \% m}$ is the coefficient of the $\lfloor{}\frac{j}{m}\rfloor{}$~th row and $j\%{}m$~th column of the response matrix, and $O_{{j-p}}$ is the $j-p$~th element of the observation / data vector. Next the relevant derivatives can be applied to \ref{eqn:decomp_eqn} to yield \ref{eqn:max_likelihood_jacobian_def} (seen at the top of the next page) where $\delta_{a, b}$ represents the Kronecker Delta function which is zero unless the subscripts match.

    With the definition of the Jacobian in hand, all we need to know is the form of the input covariance matrix, and then we can refer back to Eqn. \ref{eqn:err_with_jacobian_eqn}.
    The input covariance matrix can be defined as:
\begin{equation}
    \label{eqn:input_cov_mat}
    \Sigma^{\vec{x}}_{a,b} = \delta_{a, b}\times{}\begin{cases}
        \Delta R^2_{\lfloor \frac{a}{m} \rfloor , a \% m} & 0 \le{} a < p \\
        \Delta O^2_{a-p} & p \le{} a < p + m
    \end{cases}
\end{equation}
    Here $\Delta R_{\lfloor{}\frac{a}{m}\rfloor{},a \% m}$ is the uncertainty in the coefficient of the $\lfloor{}\frac{a}{m}\rfloor{}$~th row and $a\%{}m$~th column of the response matrix and $\Delta O_{{a-p}}$ is the $a-p$~th element of the observation / data vector.
    This definition assumes that the coefficients of the response matrix and the bins of the observed spectrum are independent.
    Without this approximation the input covariance matrix, whose \emph{full} size contains $(n+1)^2 \cdot m^2$ elements ($m$ and $n$ having the same meanings as in Eqn. \ref{eqn:variable_naming}), becomes much too large to handle.
    For even the relatively modest response matrices and spectra used in this work, such a matrix would consume a prohibitive quantity of memory (${\scriptstyle \sim}1.85$ TB.)

    The inclusion of response matrix errors increased the size of the errors of the decomposed weights by $0.1$\% to $0.2$\%.
    While this is a slight increase, it may not be the case for other uses of the decomposition method.
    In many situations, computer time is scarce, or the individual events are more computationally expensive to simulate.
    In those cases, propagation of response matrix errors will be vitally important to capture the total errors of the decomposition weights.

\subsection{Error Sources}
\subsubsection{Errors in the Observed Spectrum}
    In this work, the observed spectrum errors prior to rebinning were assumed to follow Poisson statistics and were set to the square root of the number of counts in the bin.
    The errors in the rebinned spectrum were obtained by propagating the errors of the raw spectrum through the rebinning process.
    For example a new spectrum bin that is a sum of the fractions of two old spectrum bins (as described in section \ref{sec:analysis-time-int}) the error in that bin is: $\Delta{}n = \sqrt{f_1^2 \cdot \Delta{}o_k^2 + f_2^2 \cdot \Delta{}o_{k+1}^2}$.
    Here $\Delta{}n$ is the error in the number of counts in the new bin, $f_i$ are the fractions of each of the old bins being summed into the new bin, and $\Delta{}o_j$ are the errors in the number of counts of the old raw bins.

\begin{figure*}[t!]
    % This double column equation is defined here because it appears in the next page
    % ensure that we have normalsize text
    \normalsize
    % Store the current equation number.
    \setcounter{DblColEqnCounterTemp}{\value{equation}}
    % NOTE if equations are added before this then this line needs to be updated
    \setcounter{equation}{11}
    \begin{equation}
        \label{eqn:max_likelihood_jacobian_def}
        J^{s+1}_{\mu , j} = \frac{\partial I^{s+1}_{\mu}}{\partial x_j}  = \frac{1}{\sum_{l}R_{\mu, l}} \times \begin{cases}
            -\delta_{\mu,\lfloor j/m \rfloor }I^{s+1}_{\mu} + \frac{I^s_\mu O_{j\%m} }{\sum_{\alpha} R_{\alpha, j\%m}I^s_\alpha} ( \delta_{\mu,\lfloor j/m \rfloor } - \frac{R_{\mu , j\%m}I^s_{\lfloor j/m \rfloor}}{\sum_{\alpha} R_{\alpha, j\%m}I^s_\alpha} )& 0 \le j < p \\
            &\\
            \frac{ I^s_\mu R_{\mu,j-p}}{\sum_{\alpha} R_{\alpha , j-p}I^s_\alpha} & p \le{} j < p + m
        \end{cases}
    \end{equation}
    \setcounter{equation}{\value{DblColEqnCounterTemp}}
    % The IEEE uses as a separator
    \hrulefill
\end{figure*}

% now we add 1 to the counter here to account for our double column equation above (in a weird place to force its semi ok placement in the text)
\addtocounter{equation}{1}

\subsubsection{Errors in the Response Matrix}
    In this work, errors in the response matrix coefficients were estimated from the number of counts in each bin of the energy deposition bins derived from simulation.
    Those errors were propagated through every process stage to derive the response matrix, beginning with the conversion to probability.
    Subsequently, those errors were propagated through the \emph{post facto} efficiency scaling, incorporating the estimated error in probability as well as the errors in the fitted parameters of \ref{eqn:emprical_eff_cor}.
    Finally, the errors of the scaled probability are propagated through the width convolution incorporating the errors in the fitted parameters of the peak width function.
    function.

%% file: Maximum_Likelihood_Spectrum_Decomposition_for_Isotope_Identification_and_Quantification.bbl
\begin{thebibliography}{10}
\providecommand{\url}[1]{#1}
\csname url@samestyle\endcsname
\providecommand{\newblock}{\relax}
\providecommand{\bibinfo}[2]{#2}
\providecommand{\BIBentrySTDinterwordspacing}{\spaceskip=0pt\relax}
\providecommand{\BIBentryALTinterwordstretchfactor}{4}
\providecommand{\BIBentryALTinterwordspacing}{\spaceskip=\fontdimen2\font plus
\BIBentryALTinterwordstretchfactor\fontdimen3\font minus
  \fontdimen4\font\relax}
\providecommand{\BIBforeignlanguage}[2]{{%
\expandafter\ifx\csname l@#1\endcsname\relax
\typeout{** WARNING: IEEEtran.bst: No hyphenation pattern has been}%
\typeout{** loaded for the language `#1'. Using the pattern for}%
\typeout{** the default language instead.}%
\else
\language=\csname l@#1\endcsname
\fi
#2}}
\providecommand{\BIBdecl}{\relax}
\BIBdecl
\renewcommand{\BIBentryALTinterwordstretchfactor}{4}

\bibitem{NRC_effluent}
\BIBentryALTinterwordspacing
{Nuclear Regulatory Commission}, ``{10 C.F.R. \S 50.34a Design} objectives for
  equipment to control releases of radioactive material in effluents—nuclear
  power reactors,'' accessed: 2021-03-03. [Online]. Available:
  \url{https://www.nrc.gov/reading-rm/doc-collections/cfr/part050/part050-0034a.html}
\BIBentrySTDinterwordspacing

\bibitem{DECONNINCK201361}
\BIBentryALTinterwordspacing
B.~Deconninck and C.~{De Lellis}, ``High resolution monitoring system for ire
  stack releases,'' \emph{Journal of Environmental Radioactivity}, vol. 125,
  pp. 61--68, 2013, 6th International Symposium on In Situ Nuclear Metrology as
  a Tool for Radioecology (INSINUME 2012). [Online]. Available:
  \url{https://www.sciencedirect.com/science/article/pii/S0265931X13000258}
\BIBentrySTDinterwordspacing

\bibitem{EPA_effluent}
\BIBentryALTinterwordspacing
{Environmental Protection Agency}, ``{40 C.F.R. \S 61.H, National Emission
  Standards for Emission of Radionuclides other than Radon from Department of
  Energy Facilities.}'' accessed: 2021-03-03. [Online]. Available:
  \url{https://www.govinfo.gov/content/pkg/CFR-2017-title40-vol10/xml/CFR-2017-title40-vol10-part61.xml}
\BIBentrySTDinterwordspacing

\bibitem{ortecGammaVision}
\BIBentryALTinterwordspacing
{Advanced Measurement Technology}, ``Gammavision gamma spectroscopy,''
  accessed: 2021-06-08. [Online]. Available:
  \url{https://www.ortec-online.com/products/application-software/gammavision}
\BIBentrySTDinterwordspacing

\bibitem{lucyAstro1974}
L.~B. Lucy, ``An iterative technique for the rectification of observed
  distributions,'' \emph{The Astronomical Journal}, vol.~79, p. 745, Jun. 1974.

\bibitem{dempster1977}
\BIBentryALTinterwordspacing
A.~P. Dempster, N.~M. Laird, and D.~B. Rubin, ``Maximum likelihood from
  incomplete data via the em algorithm,'' \emph{Journal of the Royal
  Statistical Society. Series B (Methodological)}, vol.~39, no.~1, pp. 1--38,
  1977. [Online]. Available: \url{http://www.jstor.org/stable/2984875}
\BIBentrySTDinterwordspacing

\bibitem{tomography1}
L.~A. Shepp and Y.~Vardi, ``Maximum likelihood reconstruction for emission
  tomography,'' \emph{IEEE Transactions on Medical Imaging}, vol.~1, no.~2, pp.
  113--122, 1982.

\bibitem{tomography2}
K.~L. Lange and R.~E. Carson, ``Em reconstruction algorithms for emission and
  transmission tomography.'' \emph{Journal of computer assisted tomography},
  vol. 8 2, pp. 306--316, 1984.

\bibitem{tainOtt2007}
\BIBentryALTinterwordspacing
J.~Tain and D.~Cano-Ott, ``Algorithms for the analysis of $\beta{}$-decay total
  absorption spectra,'' \emph{NIMA}, vol. 571, no.~3, pp. 728 -- 738, 2007.
  [Online]. Available:
  \url{http://www.sciencedirect.com/science/article/pii/S0168900206018985}
\BIBentrySTDinterwordspacing

\bibitem{rascoMTAS}
\BIBentryALTinterwordspacing
B.~C. Rasco \emph{et~al.}, ``Decays of the three top contributors to the
  reactor ${\overline{\ensuremath{\nu}}}_{e}$ high-energy spectrum,
  $^{92}\mathrm{Rb}$, $^{96\mathrm{gs}}\mathrm{Y}$, and $^{142}\mathrm{Cs}$,
  studied with total absorption spectroscopy,'' \emph{Phys. Rev. Lett.}, vol.
  117, p. 092501, Aug 2016. [Online]. Available:
  \url{https://link.aps.org/doi/10.1103/PhysRevLett.117.092501}
\BIBentrySTDinterwordspacing

\bibitem{Geant4Ref1}
\BIBentryALTinterwordspacing
S.~Agostinelli \emph{et~al.}, ``Geant4—a simulation toolkit,'' \emph{Nuclear
  Instruments and Methods in Physics Research Section A: Accelerators,
  Spectrometers, Detectors and Associated Equipment}, vol. 506, no.~3, pp. 250
  -- 303, 2003. [Online]. Available:
  \url{http://www.sciencedirect.com/science/article/pii/S0168900203013688}
\BIBentrySTDinterwordspacing

\bibitem{Geant4Ref2}
J.~Allison \emph{et~al.}, ``Geant4 developments and applications,'' \emph{IEEE
  Transactions on Nuclear Science}, vol.~53, no.~1, pp. 270--278, 2006.

\bibitem{Geant4Ref3}
\BIBentryALTinterwordspacing
J.~Allison \emph{et~al.}, ``Recent developments in {Geant4},'' \emph{Nuclear
  Instruments and Methods in Physics Research Section A: Accelerators,
  Spectrometers, Detectors and Associated Equipment}, vol. 835, pp. 186 -- 225,
  2016. [Online]. Available:
  \url{http://www.sciencedirect.com/science/article/pii/S0168900216306957}
\BIBentrySTDinterwordspacing

\bibitem{ensdfDbReport}
C.~Dunford and T.~Burrows, ``{Online Nuclear Data Service},'' International
  Atomic Energy Agency, Vienna, Austria, Tech. Rep., 1995.

\bibitem{mirionSegeDet}
\BIBentryALTinterwordspacing
{Mirion Technologies, Inc}, ``Sege\textsuperscript{TM} standard electrode
  coaxial ge detectors,'' accessed: 2021-05-28. [Online]. Available:
  \url{https://www.mirion.com/products/sege-standard-electrode-coaxial-ge-detectors}
\BIBentrySTDinterwordspacing

\bibitem{beakerGAMA}
\BIBentryALTinterwordspacing
{GA-MA \& Associates, Inc.}, ``1 liter marinelli gas containers for gas
  samples,'' accessed: 2021-05-28. [Online]. Available:
  \url{http://www.ga-maassociates.com/products/gas-beakers/1-liter-marinelli-gas-beakers/}
\BIBentrySTDinterwordspacing

\bibitem{sourceEZ}
\BIBentryALTinterwordspacing
{Eckert \& Ziegler}, ``Geometry sources - marinelli beakers (including empty
  beakers),'' accessed: 2021-05-28. [Online]. Available:
  \url{https://www.ezag.com/home/products/isotope_products/isotrak_calibration_sources/reference_sources/geometry_sources_beakers_bottles_filters_rods/marinelli_beaker_incl_empty_beakers/}
\BIBentrySTDinterwordspacing

\bibitem{GdmlRef1}
R.~{Chytracek} \emph{et~al.}, ``Geometry description markup language for
  physics simulation and analysis applications,'' \emph{IEEE Transactions on
  Nuclear Science}, vol.~53, no.~5, pp. 2892--2896, 2006.

\bibitem{livermorePhysics}
\BIBentryALTinterwordspacing
G.~Cirrone \emph{et~al.}, ``Validation of the {Geant4} electromagnetic photon
  cross-sections for elements and compounds,'' \emph{Nuclear Instruments and
  Methods in Physics Research Section A: Accelerators, Spectrometers, Detectors
  and Associated Equipment}, vol. 618, no.~1, pp. 315--322, 2010. [Online].
  Available:
  \url{https://www.sciencedirect.com/science/article/pii/S0168900210003682}
\BIBentrySTDinterwordspacing

\bibitem{livermorePhysicsImprovements}
\BIBentryALTinterwordspacing
V.~Ivantchenko, ``Recent upgrade and status of {Geant4} electromagnetic
  physics,'' in \emph{9th Geant4 Space Users' Workshop}, Mar 2013. [Online].
  Available: \url{https://indico.esa.int/event/19/contributions/1766/}
\BIBentrySTDinterwordspacing

\bibitem{convolutionTheorem}
C.~D. McGillem and G.~R. Cooper, \emph{Continuous and discrete signal and
  system analysis}, 2nd~ed.\hskip 1em plus 0.5em minus 0.4em\relax Thomson
  Learning, 1983.

\bibitem{hdf5}
S.~Koranne, ``Hierarchical data format 5: Hdf5,'' in \emph{Handbook of Open
  Source Tools}.\hskip 1em plus 0.5em minus 0.4em\relax Springer, 2011, pp.
  191--200.

\bibitem{ortecDSPECLF}
\BIBentryALTinterwordspacing
{Advanced Measurement Technology}, ``Dspec lf digital signal processing gamma
  ray spectrometer,'' accessed: 2021-06-08. [Online]. Available:
  \url{https://www.ortec-online.com/products/electronics/multichannel-analyzers-mca/workstation/dspec-lf}
\BIBentrySTDinterwordspacing

\bibitem{ortecDSPECPRO}
\BIBentryALTinterwordspacing
{Advanced Measurement Technology}, ``Dspec pro digital signal processing gamma
  ray spectrometer,'' accessed: 2021-06-08. [Online]. Available:
  \url{https://www.ortec-online.com/products/electronics/multichannel-analyzers-mca/workstation/dspec-pro}
\BIBentrySTDinterwordspacing

\bibitem{sqlite2020hipp}
\BIBentryALTinterwordspacing
D.~R. Hipp, ``Sqlite,'' accessed: 2021-06-02. [Online]. Available:
  \url{https://www.sqlite.org/index.html}
\BIBentrySTDinterwordspacing

\bibitem{hpgeHfirEffluent}
\BIBentryALTinterwordspacing
J.~Ghawaly \emph{et~al.}, ``High-purity germanium effluents measurements at the
  high flux isotope reactor and radiochemical engineering development center,''
  8 2019. [Online]. Available: \url{https://www.osti.gov/biblio/1649573}
\BIBentrySTDinterwordspacing

\bibitem{PythonTR}
G.~van Rossum, ``Python tutorial,'' Centrum voor Wiskunde en Informatica (CWI),
  Amsterdam, Tech. Rep. CS-R9526, May 1995.

\bibitem{knollRadDetAndMeasurement}
G.~F. Knoll, \emph{18.IV.B Multichannel Pulse Analysis - Spectrum Stabilization
  and Relocation - Spectrum Alignment}.\hskip 1em plus 0.5em minus 0.4em\relax
  John Wiley \& Sons, Inc., 2000, p. 702–704.

\bibitem{eigenweb}
G.~Guennebaud, B.~Jacob \emph{et~al.}, ``Eigen v3,''
  http://eigen.tuxfamily.org, 2010.

\end{thebibliography}
